\documentclass[12pt,longbibliography]{revtex4-2}
 \usepackage{graphicx,graphics,color}

\usepackage{amssymb}
\usepackage{amsfonts}
\usepackage{epstopdf}
\usepackage{bm}
\usepackage{times,xspace}
\usepackage{color}
\usepackage[utf8]{inputenc}
\usepackage[normalem]{ulem}
\usepackage{xcolor}
\usepackage{multirow}
\usepackage{physics}
\usepackage{tensor}
\usepackage{siunitx}
\usepackage{lineno}
\usepackage{amsmath}

\usepackage{etoolbox}

\newcommand*\linenomathpatch[1]{
  \cspreto{#1}{\linenomath}
  \cspreto{#1*}{\linenomath}
  \csappto{end#1}{\endlinenomath}
  \csappto{end#1*}{\endlinenomath}
}

\linenomathpatch{equation}
\linenomathpatch{gather}
\linenomathpatch{multline}
\linenomathpatch{align}
\linenomathpatch{alignat}
\linenomathpatch{flalign}

 \newcommand{\EDF}{Extended Data Fig.}
 
\newcommand{\MM}{\mathcal{M}}

\newcommand{\VV}{\mathcal{V}}

\newcommand{\tqx}{\tilde k_x}
\newcommand{\tqy}{\tilde k_y}

\newcommand{\pressure}{\raisebox{-0.25\height}{\includegraphics{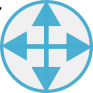}}}

\newcommand{\strone}{\raisebox{-0.25\height}{\includegraphics{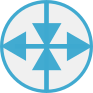}}}
\newcommand{\strtwo}{\raisebox{-0.25\height}{\includegraphics{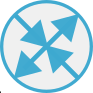}}}

\newcommand{\dilation}{\raisebox{-0.2\height}{\includegraphics{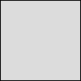}}}

\newcommand{\uone}{\raisebox{-0.1\height}{\includegraphics{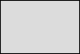}}}
\newcommand{\utwo}{\raisebox{-0.2\height}{\includegraphics{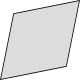}}}

\begin{document}

\title{
Limit cycles turn active matter into robots
}

\author
{Martin Brandenbourger,$^{1,2,3,\ast}$ Colin Scheibner,$^{4,5,\ast}$, Jonas Veenstra$^{1}$,\\ Vincenzo Vitelli$^{4,5,6, \dagger}$, Corentin Coulais$^{1, \ddagger}$\\
\normalsize{$^{1}$Institute of Physics, University of Amsterdam, Amsterdam, Netherlands}
\\
\normalsize{$^{2}$ TIPs Department, Universit\'e Libre de Bruxelles, Brussels, Belgium}\\
\normalsize{$^{3}$ Institut M\'ecanique et Ing\'enierie, Aix-Marseille Universit\'e, Marseille, France}\\
\normalsize{$^{4}$James Franck Institute, The University of Chicago, Chicago, Illinois 60637, USA}\\
\normalsize{$^{5}$Department of Physics, The University of Chicago, Chicago, Illinois 60637, USA}\\
\normalsize{$^{6}$Kadanoff Center for Theoretical Physics, The University of Chicago, Chicago,}
\\ \normalsize{Illinois 60637, USA}\\
\normalsize{$^\ast$ These two authors contributed equally}\\
\normalsize{$^\dagger$E-mail:  vitelli@uchicago.edu}\\
\normalsize{$^\ddagger$E-mail:  coulais@uva.nl}\\
}

\date{\today}

\baselineskip24pt

\maketitle

\medskip

\medskip

\textbf{
Active matter composed of energy-generating microscopic constituents is a promising  
platform to create autonomous functional materials~\cite{Goldman_reviewrobophysics,Aguilar_Science2018,Savoie2019robot,Palacci_Science2013,Aubret2021Metamachines,Miskin_Nature2020,tan2021development,Sanchez_Nature2012,Needleman2017,Bricard_Nature2013,Bililign2021Motile,Review_vincenzo,Marchetti2013Hydrodynamics,Ball_MRS2021}.
However, the very presence of these microscopic energy sources is what makes active matter prone to dynamical instabilities~\cite{Marchetti2013Hydrodynamics,Sanchez_Nature2012} and hence hard to control. 
Here, we show that these instabilities can be coaxed into work-generating limit cycles that turn active matter into robots. 
We illustrate this general principle in odd active media~\cite{Brandenbourger_NatComm2019,Scheibner_NatPhys2020, Bililign2021Motile, tan2021development}, model systems whose interaction forces are as simple as textbook molecular bonds~\cite{Rafsanjani_SciRob2018,Chen_PNAS2018,Liu_SciRob2018,Deng_SciAdv2020} yet not constrained to be the gradient of a potential.
These emergent robotic functionalities are 
demonstrated 
by revisiting what is arguably the oldest of inventions: the wheel. 
Unlike common wheels that are driven by external torques, an odd wheel undergoes work-generating limit cycles that allow it to roll autonomously uphill by virtue of its own deformation, as demonstrated by our prototypes.
Similarly, familiar scattering phenomena, like a ball bouncing off a wall, turn into basic robotic manipulations when either the ball or the wall is odd. 
Using continuum mechanics,
we reveal collective robotic mechanisms that steer the outcome of collisions or influence the absorption of impacts in experiments. 
Beyond robotics ~\cite{Goldman_reviewrobophysics,Correll_ReviewScience2015,Rubenstein2014Programmable,Werfel_Science2014,Aguilar_Science2018,Miskin2017graphene,Savoie2019robot,Li_Nature2019,Miskin_Nature2020,Chvykov2021low,Oliveri_PNAS2021,Aubret2021Metamachines}, work-generating limit cycles 
can also control the non-linear dynamics of active soft materials~\cite{Sanchez_Nature2012,Bricard_Nature2013,Palacci_Science2013,Keber_Science2014,Giomi_PRL2014,Aubret2021Metamachines,Bililign2021Motile,Baconnier_arxiv2021}, biological systems~\cite{Peyret_BiophysJ2019,tan2021development,Gilpin2020multiscale,Marder2001CPG, Lavi2016Cell} and driven nanomechanical devices~\cite{Miskin_Nature2020,McDonald_NatComm2020,Mathew_Nat_Nanotech2020,Miskin2017graphene}.
}

We start by asking a simple question: what is a robot?
Here, 
we take the following perspective:  
an autonomous robot---one that follows its own internal rules without receiving an external set of instructions---can be viewed as an autonomous dynamical system---one in which the time variable $t$ does not appear in the governing equation~\cite{strogatz2016nonlinear, Goldman_reviewrobophysics}. 
In this picture, cyclic robotic functionalities, such as climbing or locomotion, are enabled by robust limit cycles in the underlying dynamical system~\cite{Asano2018limit}. Limit cycles arising in living systems~\cite{Gilpin2020multiscale, Marder2001CPG, Lavi2016Cell,Bruckner2019Migrating}, active matter~\cite{Marchetti2013Hydrodynamics}, and robotic walkers~\cite{Asano2018limit, Conradt2003serpentine} 
can be driven by external stimuli, constant energy fluxes, or inertia. 
Here, we investigate limit cycles that emerge due to non-conservative, or odd, mechanical forces that naturally provide the work needed to perform the desired robotic functionalities.
Nonconservative (odd) matter is made of repeated components that obey simple local interactions, very much like ordinary matter composed of atoms and molecules, but with a crucial difference. 
In odd matter, the interaction forces between basic constituents are not constrained to be the gradient of a potential---hence by taking the system along a cycle work is generated~\footnote{Here, the usage of the word odd is distinct from other possible usages, such as odd as in parity violating, odd stress as in continuous media with local torques, and odd response coefficients that violate Onsager reciprocity.}.

\begin{figure*}[t!]
\centering
\hspace{0in}
\includegraphics[width=0.9\textwidth,trim=0cm 0cm 0cm 0cm]{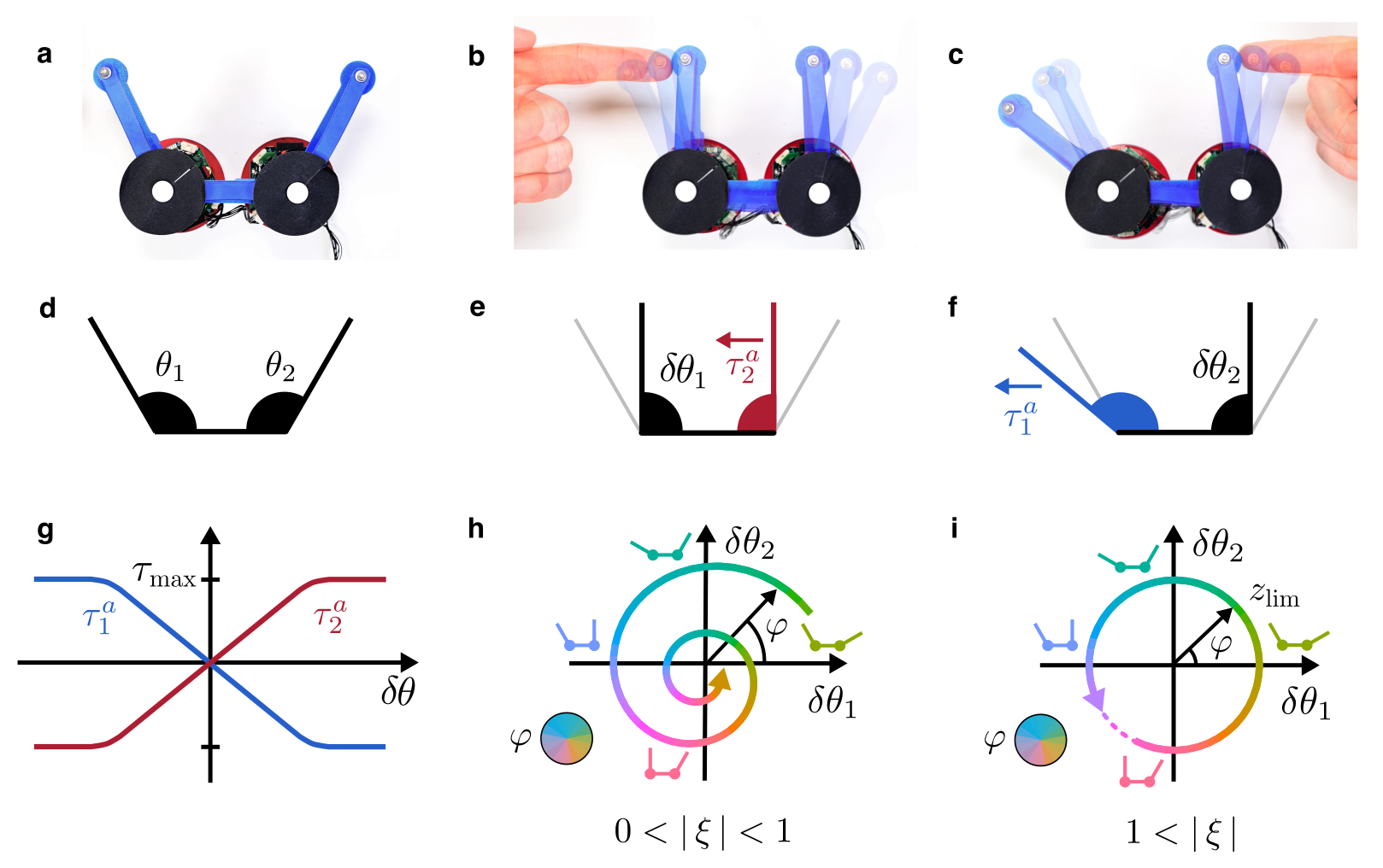}
\caption{\linespread{1.1}\selectfont{}
\textbf{Odd forces induce nonlinear work generating limit cycles} 
{\bf (a)}~Three rigid linkages are connected by motorized vertices. {\bf (b)}~A hand pushes in on the left, and the right vertex contracts. {\bf (c)}~A hand pushes in on the right, and the left vertex expands. {\bf (d-f)}~Schematics summarizing the asymmetric (or odd) stiffness for $\kappa^a>0$. 
{\bf (g)}~The nonconservative force is proportional to the angular deflection $\delta \theta$ for small amplitudes but saturates at a value $\tau_\text{max}$ at large amplitudes. {\bf (h-i)} The dynamics are parameterized by the dimensionless quantity $\xi  =  \kappa^a/ \Gamma   \sqrt{ I/\kappa } $ capturing the strength of the odd forces. For $\abs{\xi} < 1$ the system relaxes to its rest configuration. For $\abs{\xi}>1$, the odd stiffness, non-linearity, and dissipation conspire to produce a limit cycle at finite amplitude. Color indicates phase angle. See also Supplemental Video 1. The distance between two vertices is 7.5 cm. 
}
\label{Fig:block}
\end{figure*}

As an example, consider the familiar concept of bond-bending stiffness. Ordinary angular or bond-bending interactions, such as those in pendula or molecules, are approximated by a potential energy $E=\frac\kappa 2 \sum_{i} {\delta \theta_i}^2$ where the subscript $i$ labels each bond, $\delta \theta_i$ is the corresponding deviation from the equilibrium angle and $\kappa$ is the bending stiffness. The corresponding torque, obtained from gradients of the potential, is then given by the (generalized) force-displacement relation $\tau_i =-\kappa \delta \theta_i$. For a simple pendulum $\kappa$ is set by gravity.
We now revisit this familiar concept by presenting a 
minimal non-conservative generalization of bond-bending stiffness illustrated in Fig. ~\ref{Fig:block}a-c and Supplemental Video 1. Consider a system of rigid bars described by two
bond angles 
$\theta_1$ and $\theta_2$ (schematics in Fig.~\ref{Fig:block}d-f) whose force-displacement relation reads: 
\begin{align}
\mqty( 
\tau_1 \\ \tau_2
)
=\mqty( -\kappa & -\kappa^a  \\  \kappa^a & -\kappa ) 
\mqty( \delta \theta_1 \\ \delta \theta_2 )
\label{eq:micro}
\end{align}
or equivalently $\tau_i = M_{ij} \delta \theta_j$. The coefficient $\kappa$ appears in the diagonal entries of $M_{ij}$ and represents the familiar bond-bending stiffness. 
In addition to $\kappa$ the matrix $M_{ij}$ also has an off diagonal component $\kappa^a$, which represent a coupling between the two angles. 
In a passive system, e.g., two pendula coupled by a spring, the off diagonal couplings would be equal, so $M_{ij}$ would be symmetric.  
~\footnote{For this statement to apply, it is crucial that the matrix is expressed in a basis such that each component of the output vector is the conjugate force of the corresponding component of the input vector.}. 
Here, we remove this restriction and allow an antisymmetric contribution in the matrix, i.e. $M_{ij}\ne M_{ji}$. We call this contribution represented by the coefficient $\kappa^a$ odd, since it is antisymmetric under exchange of the indexes $i$ and $j$ in $M_{ij}$. 
Figure ~\ref{Fig:block} illustrates the physical meaning of this oddness: for $\kappa^a >0$ when $\theta_1$ is contracted, $\theta_2$ contracts too~(Fig.~\ref{Fig:block}b,e); when $\theta_2$ is contracted, $\theta_1$ expands instead~(Fig.~\ref{Fig:block}c,f).
The crucial point here is that this asymmetric relationship cannot exist in passive matter because the force law in Eq.~(\ref{eq:micro}) is nonconservative. When the vertices are deformed the work 
done by the vertices 
$W= \int \kappa^a (\delta\theta_2 \, \delta\dot \theta_1 - \delta\theta_1 \, \delta\dot \theta_2) \, \dd t $ depends on the path taken. Hence, if the path is cyclic, the work need not be zero; it can either be positive or negative depending on how one cycles around the loop. 
This implies that the force law in Eq. (\ref{eq:micro}) requires a source of energy to be sustained~\footnote{In the language of Hodge decomposition, such as force is an inexact form on the configuration space manifold. See the S.I. for more details on the decomposition. Moreover, we note that the single building block is free-standing in the sense that it conserves both linear and angular momentum.}.

When odd force laws such as Eq.~(\ref{eq:micro}) are combined with inertia, intuition alone suggests their dynamical fate:
any initial perturbation will grow, powered by the non-conservative forces, until non-ideal facets of the system (e.g. friction, self-collisions, or breaking) stabilize the dynamics. In some cases, the dynamical system defined by the generalized coordinates and their conjugate momenta will proceed to a limit cycle in which the energy injected by the non-conservative force balances with dissipation.
As an illustration, consider the equations of motion for the non-conservative bonds in 
Fig.~\ref{Fig:block}:
\begin{align}
     I \delta \ddot \theta_i =  - \kappa \delta \theta_i + f_i ( \delta \theta_1,  \delta \theta_2   ) - \Gamma \delta \dot \theta_i
\label{eq2}
\end{align}
where $I$ is the moment of inertia, $\Gamma$ is a dissipation coefficient, and $f_i$ is a nonlinear and nonconservative function. 
A generic form of nonlinearity, relevant also for our prototypes, is a saturation of the motors at a maximum torque $\tau_\text{max}$ (Fig.~\ref{Fig:block}h)\textemdash other geometric or dissipative sources of nonlinearities may play a similar role in non-conservative active and biological systems \cite{tan2021development,Fruchart2021non}.
Using the complex variable $z = \delta \theta_1 + i \delta \theta_ 2$ and $f= f_1 + i f_2$ we can model the nonlinear function as  
$f(z) = - i \kappa^a  z$ if $\abs{z} < \tau_\text{max}/\kappa^a$, and $f(z) = -i\tau_\text{max} z/\abs{z} $ if $\abs{z} > \tau_\text{max} /\kappa^a$, see Fig.~\ref{Fig:block}h. 
Notice that $f$ is a nonlinear extension of Eq.~(\ref{eq:micro}).
Using the ansatz 
$z = z_\text{lim} e^{-i \omega t}$, we obtain the solution~\footnote{Solving for the amplitude $\abs{z_\text{lim}}$ yields:
$\abs{z_\text{lim}} = i \tau_\text{max}/(- I \omega^2 +\kappa  - i \Gamma \omega ) $.
The constraint that $\abs{z_\text{lim}}$ must be positive and real restricts the allowed frequencies
to $\omega = -\sqrt{  \kappa/I }$. Hence, Eq.~(\ref{eq:solnon}) follows.} 
\begin{align}
    \delta \theta_1 + i \delta \theta_2 = \frac{\tau_\text{max} }{\Gamma } \sqrt{\frac I \kappa } \qty[ \cos( \sqrt{\frac \kappa I} t) + i \sin(\sqrt{\frac \kappa  I } t) ] \label{eq:solnon} 
\end{align}
up to an arbitrary phase. 
Crucially, Eq.~(\ref{eq:solnon}) is only a valid solution for sufficiently large drive. The dimensionless parameter $\xi=  \kappa^a/ \Gamma   \sqrt{ I/\kappa }$ captures the ratio of drive ($\kappa^a$) and inertia ($I$) to dissipation ($\Gamma$) and restoring forces ($\kappa$).  As shown in Fig.~\ref{Fig:block}i, when $\abs{\xi} < 1$, dissipation and restoring forces win and the process terminates at its rest configuration $z=0$. As shown in Fig.~\ref{Fig:block}j, when $\abs{\xi} > 1$, a Hopf bifurcation occurs and there exists a cycle of finite amplitude $z_\text{lim} = \tau_\text{max} /\Gamma \sqrt{ I/\kappa}$. 
Mathematically, the trajectory in Eq.~(\ref{eq:solnon}) is a limit cycle of the dynamical system in Eq.~(\ref{eq2}). 
We refer to this type of limit cycle as a nonlinear work cycle because it arises mechanically as the balance between dissipation and a non-conservative, configuration-dependent force  that performs the work 
necessary to sustain it~\footnote{In this case, the work is equal to $W =2 \pi \tau_\text{max} z_\text{lim} $}. 

We first illustrate that a robust tendency to cycle can be harnessed to perform robotic functionalities when the system is in contact with an environment (see Supplemental Video 2). 
In Fig.~\ref{Fig:wheel}a, six such vertices are connected on a closed hexagon and are levitated on an air-table (see Methods).  
In addition to solid-body translations and rotations, the hexagon has three independent deformation modes: two shear modes $S_1$ and $S_2$ and one breathing mode $B$ (Fig.~\ref{Fig:wheel}b). Due to the odd stiffness $\kappa^a$, when the hexagon is initially poked, it undergoes a spontaneous cycle of large shape change oscillating between the two shear modes (Fig.~\ref{Fig:wheel}c). 
When the table is tilted, the hexagon comes in contact with an inclined railing.
The repetitive contact provides a means of locomotion: the hexagon acts as an odd wheel. It overcomes gravity and locomotes uphill on an inclined ramp (Fig.~\ref{Fig:wheel}d) or on a rough terrain (Extended Data Fig.~\ref{Fig:Terrain}). Unlike an ordinary wheel that is driven by external torques, the odd wheel is perched on the brink of an instability that allows it to roll autonomously uphill by virtue of its own deformation. 
The dominant mechanism is an interplay between inertial rolling and internal shape change: the bottom surface of the hexagon tends to thrust left against the railing, propelling the center of mass to the right (see Fig.~\ref{Fig:wheel}e and Supplemental Video 2). This mechanism emerges from the minimal electronic feedback, rather than a directly engineered~\cite{Chiu2009deformable,Shen2006Multimode,Shen2006SuperBot,Sastra2009Dynamic,chiu2007multifunctional,Shen2002Hormone} or on-the-fly~\cite{Li_Nature2019,Oliveri_PNAS2021} locomotion algorithm. 
In contrast to limit-cycle walking based purely on inertia~\cite{Asano2018limit}, the limit cycles on display in Fig.~\ref{Fig:wheel} results from a balance of energy injection with dissipation, and hence can power uphill motion.

Beyond cyclic functionalities, such as locomotion, we show that odd matter can perform basic robotic manipulations such as steering motion and forces.  
As an example, we turn to a most basic probe of matter equally familiar to scientists or kids alike: smashing things into each other. We consider what happens when a ball collides with a wall when either the ball or the wall are odd. 
In Fig.~\ref{Fig2}ab a twelve-sided projectile with a positive odd stiffness $\kappa^a>0$ is launched vertically towards an inelastic passive surface.
Unlike a carefully swatted tennis ball, the incoming odd projectile has no initial rotation. 
Yet it deforms asymmetrically upon impact and thereby steers its outgoing trajectory to an angle determined by the sign of the odd stiffness $\kappa^a$ (Extended Data Fig. 1a and Supplemental Video 3). To make sense of this robotic functionality, we examine the transient dynamics inside the odd projectile.
In Fig.~\ref{Fig2}bc, we reconstruct the full internal degrees of freedom of the projectile and plot the angular deviation of each vertex as a function of time in the associated heat map. 
The qualitative similarity between the active and passive bending stiffness in Eq.~(\ref{eq:micro}) suggests an approach to understanding the collision process: collective modes. 
In Fig.~\ref{Fig2}d, we project the vertex dynamics onto the lowest two Fourier modes of the ring. Physically, these Fourier modes represent a shearing deformation analogous to the $S_1$ and $S_2$ deformation of the hexagon.
The out-spiraling motion in Fig.~\ref{Fig2}d captures the lowest mode contributions to the unidirectional wave propagating counterclockwise in Fig.~\ref{Fig2}c. 
This out-spiraling behavior is powered by activity and is eventually stabilized by self intersection, dissipation, and motor nonlinearities. 

We now show that treating the odd projectile as a non-conservative continuous material provides basic insights into the outcome of the impact. 
We model the chain as a continuum fiber with periodic boundary conditions (Fig.~\ref{Fig2}e-top inset). 
By linearizing the equations of motion and taking the continuum limit of a 1D chain of active vertices (see S.I.), we arrive at the following distinctive wave equation:
\begin{align} 
(\partial_\tau^2  + \partial_\lambda^4 + 2 \tilde \kappa^a \partial_\lambda^5 + \tilde \Gamma \partial_\lambda^4 \partial_\tau) h(\lambda, \tau) =0 \label{eq:wave},
\end{align}
where $h(\lambda, \tau)$ is the amplitude of the fiber's deflection as a function of time $\tau$ and space  $\lambda$, while $\tilde{\Gamma}$ is the dissipation, $\tilde \kappa^a=\kappa^a/N\kappa$, and $N$ is the number of unit cells, see Methods.  When $\kappa^a=0$, Eq.~(\ref{eq:wave}) is the standard wave equation for a thin beam with bending stiffness. Crucially, $\kappa^a\neq 0$, a new term is introduced that violates $\lambda \to -\lambda$. The broken parity and Hermiticity imply that the waves will be amplified in one direction relative to the other, an effect known as the non-Hermitian skin effect~\cite{MartinezAlvarez_PRB2018,Yao_PRL2018,Brandenbourger_NatComm2019,Ghatak_PNAS2020,Ronny_NatPhys2020,Zhong_NatPhys2020,Chen_NatComm2021,Coulais_NatPhys2020,Bergholtz_RMP2021,Review_vincenzo,Scheibner2020non,Zhou2019non}.

The unidirectional propagation is associated with shape-changes that control the bounce. 
Initially, when the projectile compresses under its own inertia, the deformation induced in the impact corresponds to the ring's second harmonic, of wavelength $L/2$. 
Starting from this initial condition, the analytical solution for the curvature $\partial_\lambda^2 h$ over time (Fig.~\ref{Fig2}e-bottom inset) captures the angular deflection at each vertex appearing in the experimental color-map of Fig.~\ref{Fig2}c.
This unidirectional wave is responsible for the asymmetry and the enhancement of the rebound. While in contact with the surface, the bumps of negative curvature (blue in Fig.~\ref{Fig2}f inset) on the sides propagate in one direction: the left bump travels down, while the right bump travels up. The downward push on the left propels the center of mass to the right and gives it an additional vertical thrust. As implied by symmetry considerations, in Fig.~\ref{Fig2}f, the vertical momentum transfer $p_{y,\text{out}}/p_{y,\text{in} }$ is even in $\kappa^a$ while the transverse momentum transfer $p_{x, \text{out}}/p_{y,\text{in} } $ is odd. 
Using a minimal model of a solid block only capable of shearing, we are able to derive analytical expressions (solid lines in Fig.~\ref{Fig2}f, see Methods) that capture the general experimental trends (red and blue markers for $\kappa^a> 0$ and $\kappa^a<0$ respectively).  

Finally, we examine an additional robotic functionality: an odd wall can steer post-impact vibrations and the outgoing trajectory of the bullet. We consider the scenario of a passive projectile striking against a two-dimensional odd wall with positive odd stiffness $\kappa^a > 0$ (Fig.~\ref{Fig3}a) and negative odd stiffness $\kappa^a < 0$ (Extended data Fig. 2). 
The odd wall consists of 17 odd hexagons (120 powered vertices in total) patterned into a honeycomb lattice (see S.I.).
Upon impact, we observe a rotation and deflection of the passive projectile
and a high degree of asymmetry of the vertical displacement field $u_y(x,y)$ in the odd solid, reconstructed from experiments in Fig.~\ref{Fig3}b. Fig.~\ref{Fig3}c shows $u_y(x,t)$ averaged in the $y$ direction, revealing that the wall 
undergoes stronger vibrations on its left edge during and after the impact (See Supplemental Video 4). In Fig.~\ref{Fig3}d, we see that the spatially averaged components ($S_1$ and $S_2$) of the shear strain in the wall exhibits a cyclic coupling reminiscent of the locomotion in Fig.~\ref{Fig:wheel}f and impact in Fig.~\ref{Fig2}d .

Since the impact absorption relies on the coordinated motion of many interacting components, we ask whether the 
emergent features can be captured in a way agnostic to microscopic details, i.e. via continuum elasticity.
Passive, isotropic 2D elasticity is summarized by just two parameters: the Young's Modulus $E$ and the Poisson's ratio $\nu$. For media with nonconservative interactions, an additional parameter exists: the odd ratio $\nu^o$~\cite{Scheibner_NatPhys2020}. The odd ratio captures the nonconservative relationship between shear stress and shear strain, see Methods. 
In the present realization, the odd ratio, a macroscopic quantity, relates to the microscopic odd stiffness as $\nu^o \approx 0.58 (\kappa^a/\kappa)$ (see S.I.). We can utilize this non-conservative continuum theory, known as odd elasticity~\cite{Scheibner_NatPhys2020}, to rationalize basic features of the internal dynamics inside the wall under impact.

We quantify the steering of the post-impact vibrations by measuring the center of vibration $\ell$ defined in Eq.~(\ref{eq:ell}) as a function of $\nu^o$ in Fig.~\ref{Fig3}e. To rationalize the dependence of $\ell$ on $\nu^o$, we consider the ideal case of a half-infinite continuum medium with $\nu^o\neq0$. In the Methods, we show that such a medium can host unidirectionally amplified Rayleigh waves (Fig.~\ref{Fig3}e-top inset). Crucially when $\nu^o \neq 0$, these waves acquires a nonzero exponential amplification along the boundary, quantified by an inverse growth length $q_x$, such that $u_y(x) \propto e^{q_x x}$ (Fig.~\ref{Fig3}e-bottom inset). This unidirectional amplification, reminiscent of the non-Hermitian skin effect~\cite{MartinezAlvarez_PRB2018,Yao_PRL2018,Brandenbourger_NatComm2019,Ghatak_PNAS2020,Ronny_NatPhys2020,Zhong_NatPhys2020,Chen_NatComm2021,Coulais_NatPhys2020,Bergholtz_RMP2021,Review_vincenzo,Scheibner2020non,Zhou2019non}, is what determines the steering of post impact vibrations, i.e., the sign of $\ell$ in Fig.~\ref{Fig3}e main panel.

To intuit how the odd wall manipulates the outward trajectory of the passive bullet, we examine the asymmetry in the displacement field of the wall at the deepest point of impact. We consider an idealized situation in which a point force is exerted on the boundary of the semi-infinite odd medium (Fig.~\ref{Fig3}f). In this case, the continuum solution for the vertical component of the static displacement field $u_y$ takes the simple form (see Methods):
\begin{align}
    u_y = \frac{2F}{\pi E} \left[ 2 \log r +1 -  \cos 2\phi + \nu^o (\sin 2 \phi +2)  \right] \label{eq:uy}
\end{align}
where $r$ and $\phi$ are polar coordinates about the point of impact, 
$F$ is the applied force, which is taken to be the mass of the projectile times its deceleration. The odd ratio, $\nu^o$ in Eq.~(\ref{eq:uy})
introduces a term that violates the $\phi \mapsto -\phi$ symmetry,  a continuum manifestation of the experimentally observed asymmetry. In Fig.~\ref{Fig3}g, we plot the outgoing angle $\alpha$ of the projectile as a function of $\nu^o$. The sign of $\alpha$ can be inferred 
qualitatively via the $S_1$-$S_2$ shear coupling at the point of contact, as sketched in Fig.~\ref{Fig3}g (inset) for a single motorized hexagon and highlighted in Fig.~\ref{Fig3}a (inset) for the two powered hexagons touching the projectile. 
Beyond the first layer, the pattern continues smoothly further away from the point of impact~Fig.~\ref{Fig3}b. 
This suggests that the discrete behavior should be evident in the continuum solution. Indeed, the sign of the slope ($\partial_x u_y\propto\nu^0$, see Fig.~\ref{Fig3}f dashed line) at the point of contact in the continuum agrees with the sign of the tilt imparted to the projectile inferred from the discrete mechanism and observed in experiment.

We have shown how macroscopic robotic functionalities can emerge from simple non-conservative interactions. 
Identifying and engineering these cycles offers an approach to tailoring functionalitiies across platforms:
the key ingredients to support these work-generating limit cycles are present in biological systems such as spinning embryos or bacteria~\cite{tan2021development,Petroff2015fast}, biomembranes~\cite{Needleman2017,Julicher2018,Banerjee2021active,Kole2021Layered}, or driven colloidal systems \cite{Bililign2021Motile,Yan2015,Grzybowski2000Dynamic,Aragones2016elasticity}, as well as micro- and nanomechanical devices~\cite{Liu2021micrometer,Miskin_Nature2020,Mathew_Nat_Nanotech2020,Miskin2017graphene}. 
In analogy with how a sequence of amino acids folds into a functional protein, we envision an organic design process in which minimal active units evolve towards a functional dynamic state represented by a limit cycle~\cite{Anupama2017cargo}. 
It is an open question how such principles generalize on scales in which thermal or environmental fluctuations are relevant and the resulting statistical mechanics is no longer based on energy minimization~\cite{Fruchart2021non,Hong2011Kuramoto}.



%


\clearpage

\clearpage 

\begin{figure*}[t!]
\centering
\hspace{0in}
\includegraphics[width=0.9\textwidth,trim=0cm 0cm 0cm 0cm]{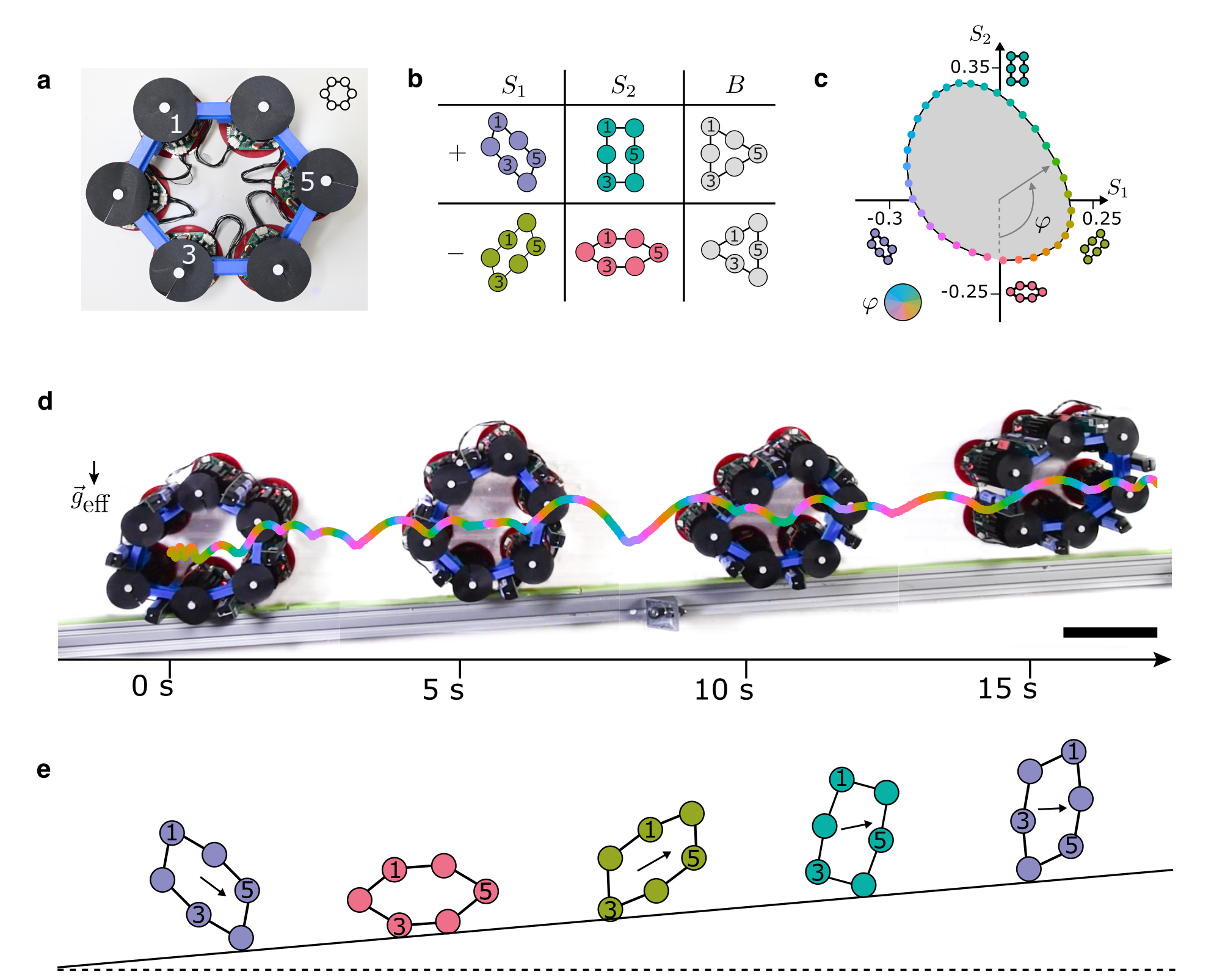}
\caption{\linespread{1.1}\selectfont{}
\textbf{Reinventing the wheel: nonlinear work generating limit cycles power locomotion}  
{\bf (a-b)} Six odd vertices are connected in a hexagon, whose shape is summarized by 3 independent angles, $\theta_1$, $\theta_3$, $\theta_5$, or alternatively three deformation modes: two shears $S_1$ and $S_2$, and a breathing mode $B$. {\bf (c)} When initially perturbed, the system evolves towards a limit cycle in the space of $S_1$ and $S_2$. Color indicates the phase angle $\varphi$.  {\bf (d)} The cycles compete with gravity to move the odd wheel up a ramp. The projection of gravity onto the air table $\vec g_\text{eff}$, points downward. See Methods for experimental details. 
{\bf (e)} A simplified schematic of the propulsion mechanism arising from the cycle. See also Supplemental Video 2.  Scale bar: 10 cm. 
}
\label{Fig:wheel}
\end{figure*}

\clearpage

\begin{figure*}[t!]
\centering
\hspace{0in}
\includegraphics[width=0.9\textwidth,trim=0cm 0cm 0cm 0cm]{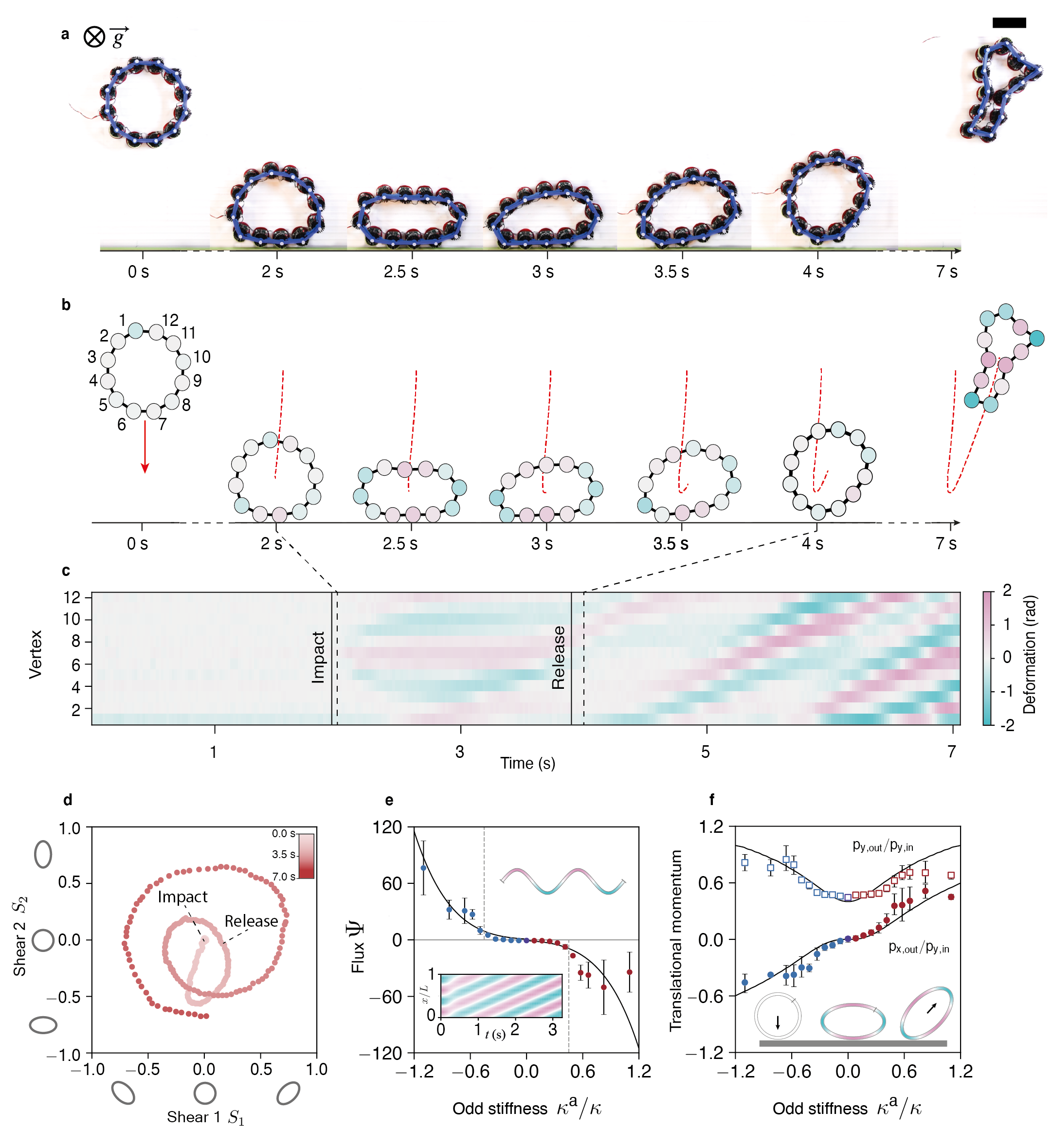}

\caption{
\linespread{1.1}\selectfont{}
\textbf{Impact of an odd projectile against passive wall} 
{\bf (a) } An odd projectile with $\kappa^a>0$ is vertically incident against a wall. Scale bar: 10 cm.
{\bf (b-c)} The angular deformation of each vertex is plotted as a function of time, revealing a unidirectional wave.
{\bf (d)} The lowest two deformation modes are plotted as a function of time.  
{\bf (e)} (Inset top) The chain is modeled as a continuous ring. Inset, bottom: Eq.~(\ref{eq:wave}) is evolved on a 1D periodic ring with the second harmonic is used as an initial condition. Color indicates deviation in curvature. Main: the flux $\Psi$ [see Methods Eq.~(\ref{eq:flux}) ] quantifies the unidirectionality and amplification of the waves in experiment. Black lines are theoretical predictions based on Eq.~(\ref{eq:wave}). The vertical dashed lines denote values of $\kappa^a/\kappa$ at which the system exhibits a bifurcation analogous to that in Fig.~\ref{Fig:block}ef, see Eq.~(\ref{eq:bifur}). {\bf  (f) } We model the projectile as a box that is only allowed to shear (see Methods). The outgoing momenta $p_x$ and $p_y$ as a function of $\kappa^a/\kappa$. Markers are experiments and the solid line is the minimal model. (Inset) A mechanical interpretation of the unidirectional wave guiding the impact. For markers in (e-f), color indicates sign of $\kappa^a$. 
See Supplemental Video 3. 
}
\label{Fig2}
\end{figure*}

\clearpage

\begin{figure*}[t!]
\centering
\hspace{0in}
\includegraphics[width=0.85\columnwidth]{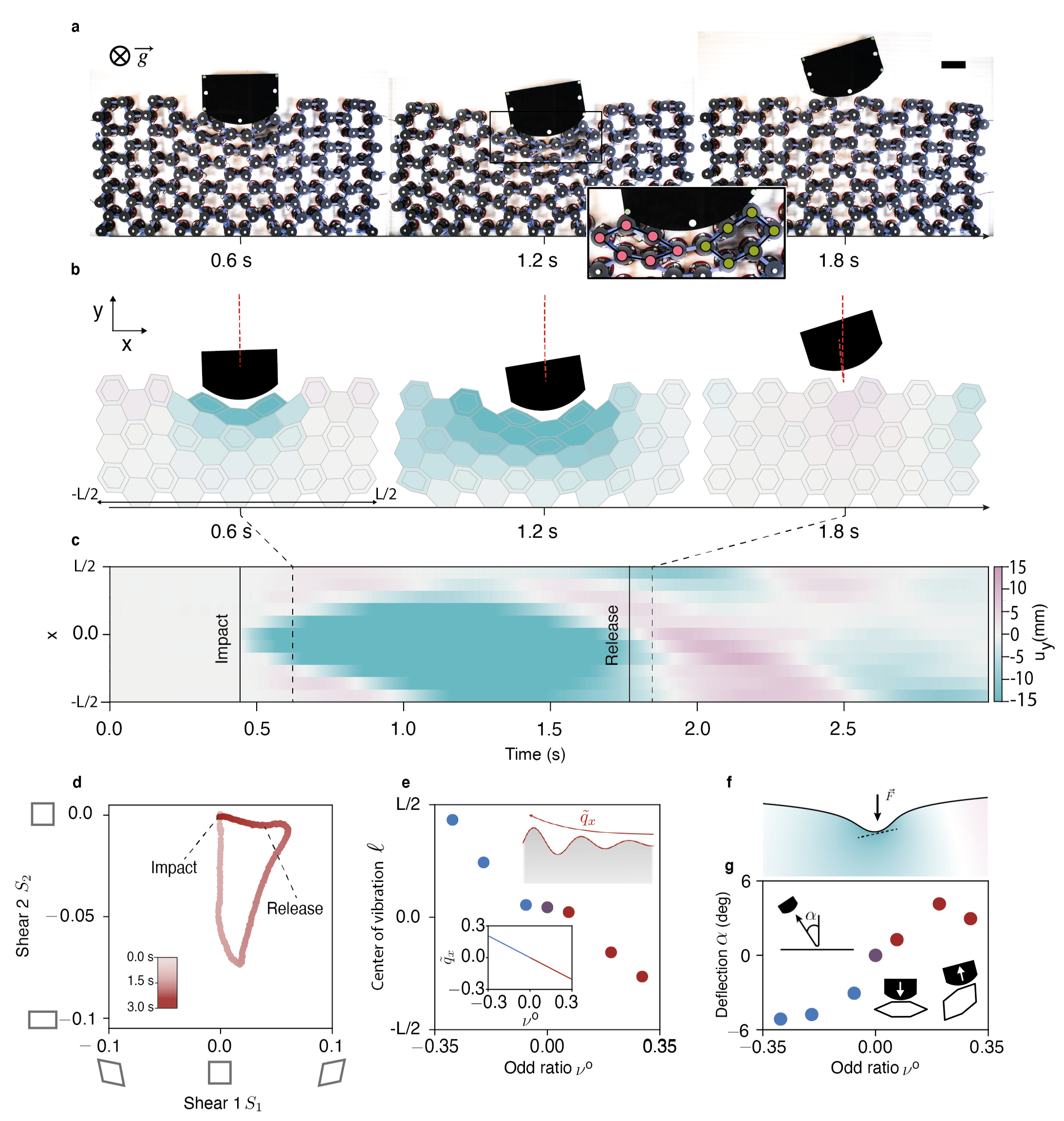}
\caption{\linespread{1.1}\selectfont{}
\textbf{Impact of a passive projectile on an odd wall} {\bf (a)} A passive projectile is launched at a wall of 120 motorized vertices ($\kappa^a >0$), forming a tiling 17 robotic hexagons [denoted by the double-line hexagons in (b).] Scale bar: 10 cm. (Inset) Active vertices colored by their deformation, cf.~Fig.~\ref{Fig:wheel}c. 
{\bf (b)} Experimental reconstructions of the wall. Color represents the average vertical displacement $u_y$ of each hexagon and L=128 cm. 
{\bf (c)} The displacement averaged over each column $u_y$ as a function of time and horizontal coordinate $x$.
{\bf (d)} The average shear deformation in the wall with color indicating time. 
{\bf (e) } The center of vibration $\ell$ [see Eq.~(\ref{eq:ell})] measured as a function of the odd ratio $\nu^o$. (Inset top right) The continuum model exhibits Rayleigh waves that grow exponentially along the surface with inverse decay length $q_x$, see Eqs.~(\ref{eq:sec1}-\ref{eq:sec2}). (Inset bottom left)  $\tilde q_x = q_x \sqrt{\frac{E}{\rho \omega^2}}$ as a function of the odd ratio $\nu^o$ predicted by the continuum theory, where $\omega$ is the frequency of the wave, and $E$ and $\rho$ are the Young's modulus and density of the wall.
{\bf  (f)} Continuum solution to a point force for $\nu^o =0.2$, echoing the asymmetric deformation of the wall observed upon impact, see Eq.~(\ref{eq:uy}).  The same color scheme as (b). The dashed line indicates the slope at the point of contact.  
{\bf (g) }
Measured angle of deflection $\alpha$ with respect to the vertical. For markers in (e-f), color indicates sign of $\kappa^a$. (Inset lower right) Deflection rationalized via a single odd hexagon under impact. 
See also Supplemental Video 4.
}
\label{Fig3}
\end{figure*}

\clearpage 

\section{Extended data Figures}

\setcounter{figure}{0}
\renewcommand{\thefigure}{\arabic{figure}}
\renewcommand{\figurename}{\EDF}

\setcounter{equation}{0}
\renewcommand{\theequation}{M\arabic{equation}}

\begin{figure*}[h!]
\centering
\hspace{0in}
\includegraphics[width=0.9\columnwidth]{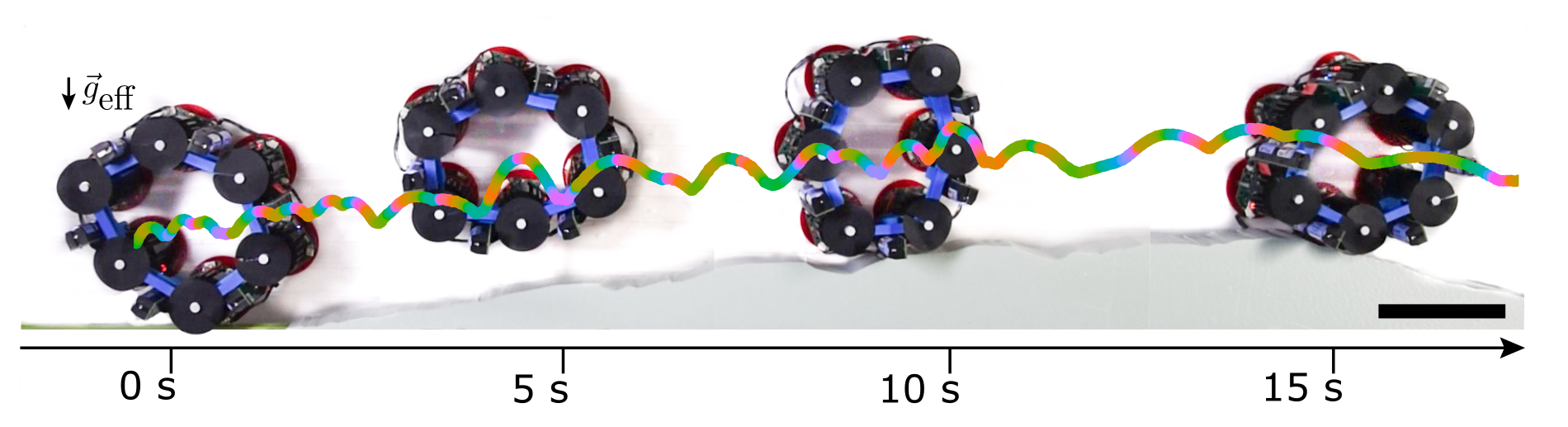}
\caption{\linespread{1.1}\selectfont{}
\textbf{Locomotion over complex terrain.} The odd wheel from Fig.~\ref{Fig:wheel} climbs over a complex terrain. The line tracks the center of mass and the color bar is the same as in Fig.~\ref{Fig:wheel}. See Methods for experimental details. See also Supplemental Video 2.
}
\label{Fig:Terrain}
\end{figure*}

\begin{figure*}[h!]
\centering
\hspace{0in}
\includegraphics[width=0.9\columnwidth]{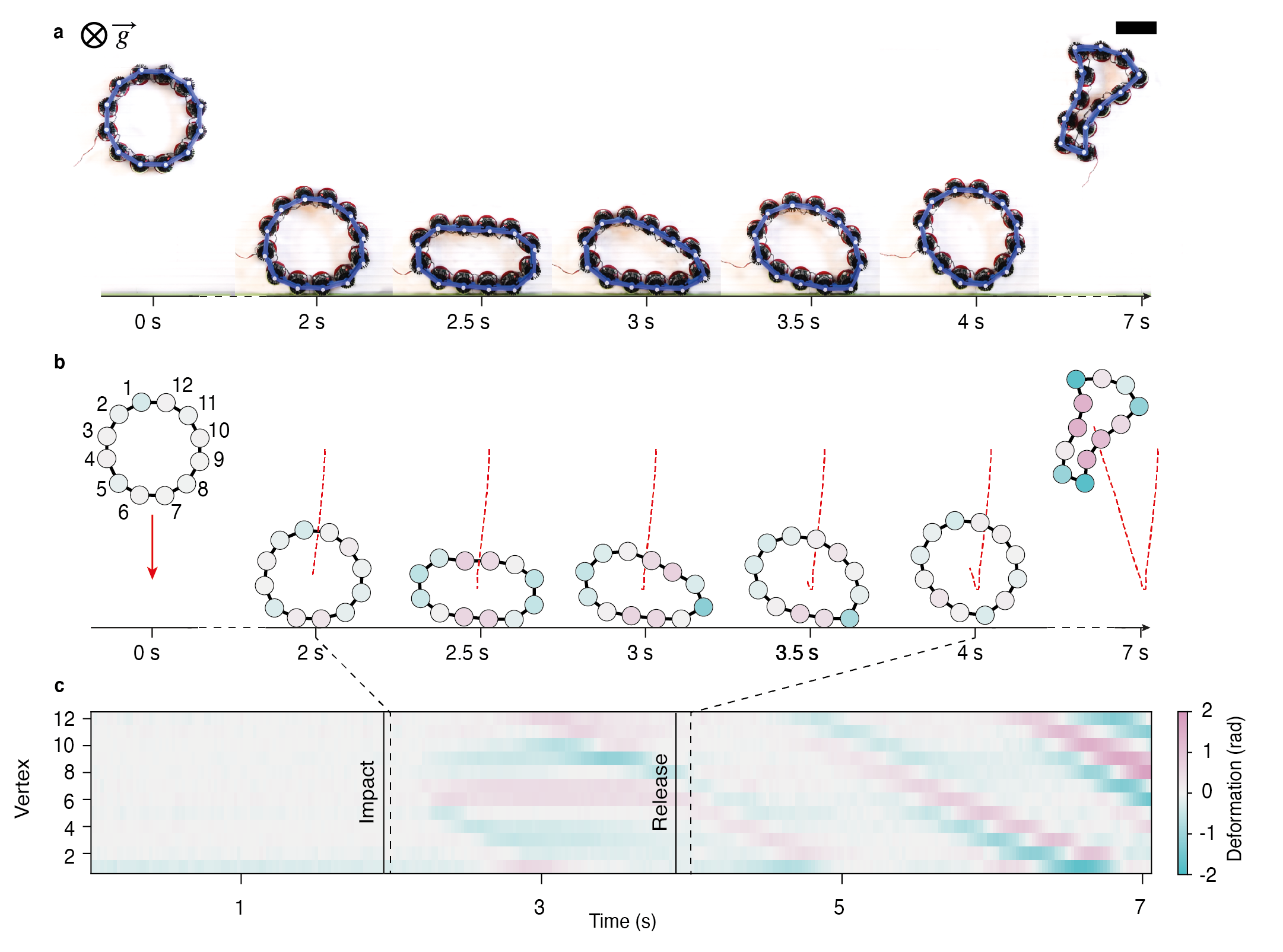}
\caption{\linespread{1.1}\selectfont{}
\textbf{Impact of an odd projectile against passive wall} 
{\bf (a) } An odd projectile with $\kappa^a<0$ is vertically incident against a wall. Scale bar: 10 cm.
{\bf (b-c)} The angular deformation of each vertex is plotted as a function of time, revealing a unidirectional wave propagating in the opposite direction of Fig. \ref{Fig2}bc. See also Supplemental Video 3. } 
\label{Fig:Odd_projectile_extended}
\end{figure*}

\begin{figure*}[h!]
\centering
\hspace{0in}
\includegraphics[width=0.9\columnwidth]{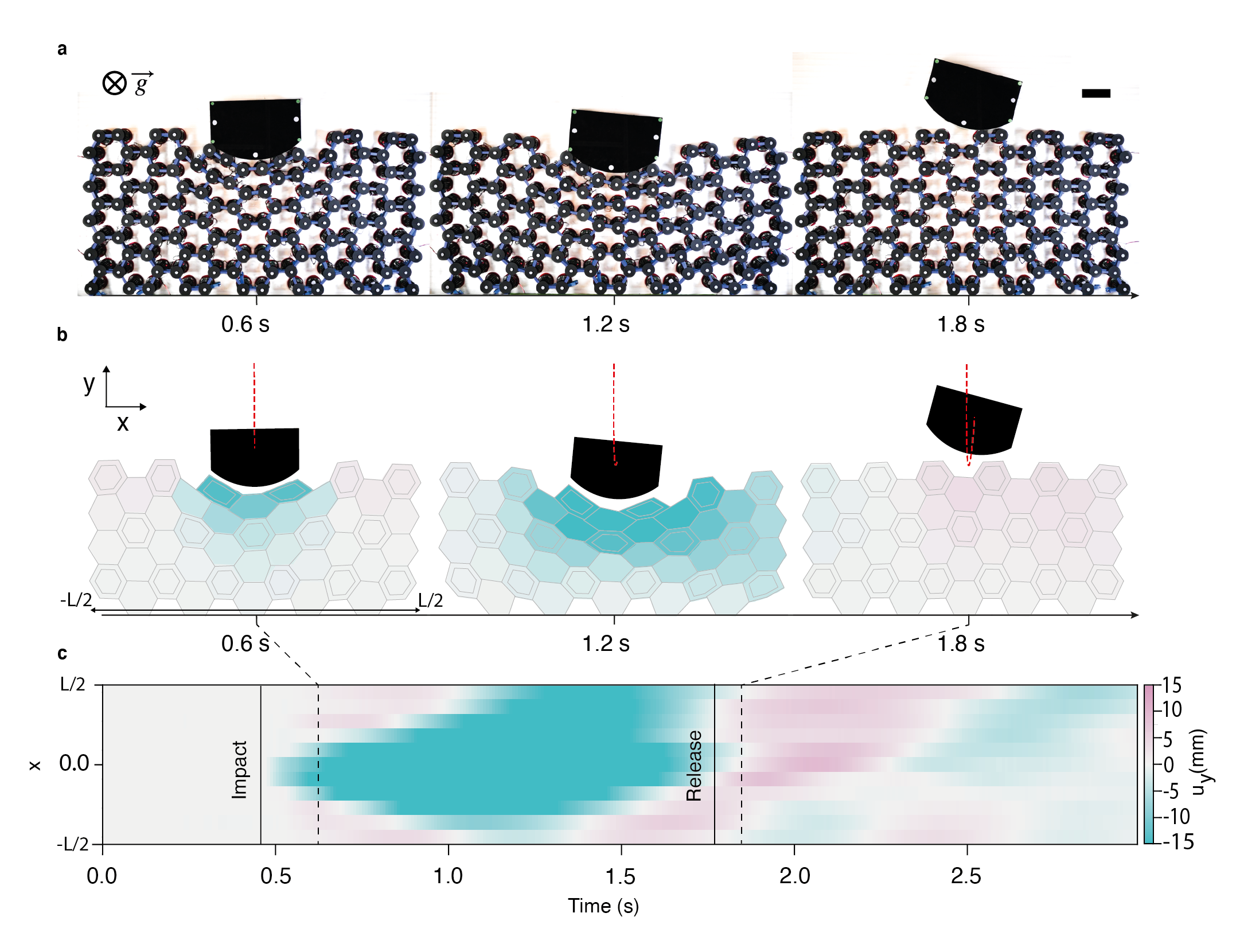}
\caption{\linespread{1.1}\selectfont{}
\textbf{Impact of a passive projectile on an odd wall} {\bf (a)} A passive projectile is launched at a wall of 120 motorized vertices with $\kappa^a <0$. Scale bar: 10 cm.
{\bf (b)} Corresponding reconstructions of the wall, where the vertical displacement $u_y$ of each hexagon is plotted. The total length is $L=128$~cm. 
{\bf (c)} The displacement is averaged over each column $u_y$ and is shown as a function of time and of the horizontal coordinate $x$. The deformation over time is opposite to the one observed in Fig. \ref{Fig3}. See also Supplemental Video 4.
}
\label{Fig:Odd_wall_extended}
\end{figure*}

\clearpage

\clearpage

\section{Materials and Methods}

\subsection{Construction of the robotic materials} \label{sec:construction}
The linkages shown in Fig.~\ref{Fig:block}a-c are composed of motorized vertices connected by plastic arms. 
Each vertex consists of a DC coreless motor (Motraxx CL1628) embedded in a cylindrical heatsink, an angular encoder (CUI AMT113S), and a microcontroller (ESP32) connected to a custom electronic board. The electronic board enables power conversion, interfacing between the sensor and motor, and communication between vertices.
Each vertex has a diameter 50 mm, height 90 mm, and mass $0.2$ kg. 
The power necessary to drive the motor is provided either by two Vapcell 16340 T8 batteries (Fig.~\ref{Fig:wheel}d) or by an external 48 V DC power source (Figs.~\ref{Fig2},\ref{Fig3}).

Rigid 3D-Printed arms connect each motor's drive shaft to the heat sink of the adjacent unit. 
The angle formed between the two arms at vertex $i$ is denoted by $\theta_i$. The on-board sensor measures $\theta_i$ at a 
sample rate of $100$ Hz and communicates the measurement to nearest neighbors. 
In response to the incoming signal, vertex $i$ exerts an active torsional force $\tau_i^a$
\begin{equation}
\tau_i^a = \kappa^a \left(\delta \theta_{i-1}-\delta \theta_{i+1} \right). \label{eq:fb} 
\end{equation}
where $\delta \theta_i = \theta_i - \theta^0$ and $\theta^0$ is the rest angle. 
The active stiffness $\kappa^a$ was programmed by the microcontroller and calibrated by measuring torque-displacement slopes for different values of the electronic feedback. Each arm has a length of $a=7.5\,\mathrm{cm}$. 
The coreless motor saturates at a maximum torque of $\tau_\mathrm{max} = 12$ mN\,m.
Adjacent vertices are also connected by rubber bands (blue in Fig.~\ref{Fig:block}a) which provide a passive elastic torsional stiffness denoted by $\kappa=48$ mN m/rad. See Section~S2 of the S.I. for calibration details.  

\subsection{Experimental protocol}

Experiments take place on top of a custom-made low friction air table. 
The table consists of two $1.5 \, \si{m}\times 1.5\,  \si{m}$  plexiglas plates that sandwich 5 mm-wide air channels. The top plate is pierced by an array of  holes (pitch 10~mm, diameter 1~mm) conveying air pressurized at 10~bars. 
Each motorized vertex floats on a thin layer of pressurized air without making contact with the table. The experiments are filmed via a Nikon D5600 camera equipped with a 50 mm lens, recording 2 Mpx images at 50 frames per second. A white marker is placed on the top of each vertex, and we use the python package OpenCV to detect and track the position of each vertex with a spatial resolution of 0.8 mm.

In Fig. \ref{Fig:wheel}a-c, a device consisting of 6 vertices (shown in Fig.~\ref{Fig:wheel}a) was subjected to a manually applied initial compression. Regardless of the exact initial perturbation, a stable cyclic motion emerged.
We parameterize the internal state via the three degrees of freedom
\begin{align}
    S_1 =& \frac1{\sqrt 2} ( \delta \theta_1 - \delta \theta_3 ) \\ 
    S_2=&  \frac1{\sqrt 5} (\delta \theta_1 + \delta \theta_3 - 2 \delta \theta_5) \\ 
    B=&  \frac1{\sqrt 3} (\delta \theta_1 + \delta \theta_3 + \delta \theta_5) 
\end{align}
The projection of experimental data onto the $S_1$, $S_2$ axis is shown in Fig.~\ref{Fig:wheel}c. 

In Fig. \ref{Fig:wheel}d and Extended Data Fig.~\ref{Fig:Terrain}, the air table was tilted by $6^\circ$  with respect to the horizontal plane. This creates an effective gravitational constant of $g_{\mathrm{eff}}\approx 1\,\text{m}\,\text{s}^{-2}$ that points towards the aluminum profile and tends to push the robotic hexagon against the profile. In turn, the aluminum profile (initially parallel to the floor) is itself rotated $5^\circ$ about the normal of the table. 
The total increase in altitude of the robotic hexagon when it climbs across the profile is thereby $1.5 \,\si{m} \times\sin 6^\circ \times\sin 5^\circ =1.4$~cm, see Supplemental Fig.~S4.
In Fig.~\ref{Fig:wheel}d, strips of vinylpolysiloxane (Elite Double 32) are placed along the aluminum profile to provide additional friction between the robotic material and the boundary.

The experiments shown in Fig.~\ref{Fig2} and Fig.~\ref{Fig3} are performed on a horizontal air table. In Fig.~\ref{Fig2}, a projectile consisting of 12 active vertices is launched towards an aluminum profile. Two initial velocities, 0.18~m/s and 0.26~m/s, are used.
In Fig.~\ref{Fig3}, a projectile of mass 8.2~kg was propelled at 0.5~m/s towards an active hexagonal lattice of vertices. 
The active feedback rule in Eq.~(\ref{eq:fb}) was implemented on vertices connected by the double line in Fig.~\ref{Fig3}b (see also Supplemental Fig.~S3). 
The bottom row of the wall is anchored to the aluminum profile to prevent slipping.  
For all experiments, the incoming projectile velocity was set by an aluminum profile connected to a conveyor belt and a stepper motor (Beckhoff AM8112 and controller EL7211-9014).  
In Fig.~\ref{Fig2} and Fig.~\ref{Fig3}, the value of $\kappa^a$ assumes values in the range $[-22, 22]$~mN~m/rad. As detailed in Section~S2 of the S.I., simulations are used to determine the odd ratio $\nu^0$ as a function of $\kappa^a/\kappa$.

\subsection{Active projectile: wave mechanics and center of mass motion}

Here we detail the analysis of the active projectile under impact. When the projectile makes contact with the wall, it is deformed under its own inertia into an elliptical shape, which represents a shearing of the enclosed 2D area. We model the collision by treating the elliptical deformation as as a second harmonic distortion of a periodic ring. 

\subsubsection{Non-Hermitian wave mechanics}

In Section~S1 of the S.I., we provide the nonlinear equations of motion for an idealized 1D chain of vertices of arbitrary shape. The simplest reduction of this nonlinear equation is to assume that the chain is initially straight, has periodic boundaries, and total length $L$.
The linearized continuum equation of motion to order $a^3$ for the chain is given by
\begin{align}
    m \partial_t^2 h + a^2 \kappa \partial_x^4 h + 2 a^3 \kappa^a \partial_x^5 h + a^2 \Gamma  \partial_x^4 \partial_t h =0 \label{eq:pde}
\end{align}
where $h$ is the transverse deflection of the vertices.  
It is useful to write the equations in dimensionless  form:
\begin{align}
     \partial_\tau^2 h + \partial_\lambda^4 h + 2 \tilde \kappa^a \partial_\lambda^5 h + \tilde \Gamma \partial_\lambda^4 \partial_\tau h =0 \label{eq:pde_dim}
\end{align}
where we have introduced the quantities 
\begin{align}
\tilde \kappa^a \equiv& \frac{\kappa^a}{N \kappa} \\ 
\tilde \Gamma \equiv & \frac{\Gamma}{N^2 a \sqrt{\kappa m} } \\  
\omega_0 \equiv& \frac{1}{N^2} \sqrt{ \frac {\kappa}{m a^2}} \\  
\tau \equiv& t \omega_0 \\ 
\lambda \equiv& \frac{x}{L} 
\end{align}
where $N = L/a$ is the number of unit cells. 
From Eq.~(\ref{eq:pde_dim}), the dispersion is given by:
\begin{align}
\tilde \omega_\pm = - \frac{i \tilde \Gamma}2 \tilde q^4 \pm \sqrt{ - \left( \frac {\tilde \Gamma}{ 2} \right)^2 \tilde q^8 +\tilde q^4 +2 i\tilde \kappa^a \tilde q^5} \label{eq:dispersion} 
\end{align}
where $\tilde \omega \equiv \omega/\omega_0$ is the dimensionless frequency and $\tilde q = qL$ is the dimensionless wave number. In the experiment, $N=12$, $a=7.5$ cm, $m=0.2$ kg, and $\kappa^a$ is the control parameter. We use $\kappa = 20$ mN m/rad and $\Gamma = 4$ mN m s/rad, determined as fitting parameters between theory and experiment.

In Fig.~\ref{Fig2}a and \EDF~\ref{Fig:Odd_projectile_extended}, we show runs of the experiment for $\kappa^a=\pm 9.9$~mN m/rad. The heat maps show the angular deviation $\delta \theta_i$ for each vertex as a function of time. In the continuum model, the curvature  $ c(x) = a^2 \partial_x^2 h$ is the coarse-grained variable corresponding to the local angular deflection $\delta \theta_i$. In inset of Fig.~\ref{Fig2}e, we show the curvature distribution corresponding to the solution: 
\begin{align}
 h(\vb x, t) = A \left[ e^{i ( \frac{4 \pi  x}L - \omega_+ t ) } + e^{i ( \frac{4 \pi x}L - \omega_- t)} + \text{c.c.}  \right ]
\end{align}
where $A$ is an overall amplitude and $\text{c.c.}$ stands for complex conjugate, and $\omega$ is evaluated at $q = 4\pi/L$. In Fig.~\ref{Fig2}e,
we provide vertical, gray dashed lines where the imaginary part of the spectrum 
\begin{align}
\omega_\textrm{Im}= \omega_0 \cdot  \bigg[ \textrm{Im} (\tilde \omega) \bigg]_{\tilde q = \textrm{sgn}(\tilde \kappa^a) \cdot 4 \pi} \label{eq:bifur}
\end{align}
passes from positive to negative. When $\omega_\text{Im} > 0$ ($\omega_\text{Im} <0$), the  $\tilde q = 4 \pi$ mode is amplified (attenuated). 

To quantitatively compare between model and experiment, we compute the real and imaginary parts of the lowest harmonic modes from the experimental data:
\begin{align}
    S_1 (t) =& -\frac16 \sum_{j =1}^{12} \cos ( j \pi/3)  \delta \theta_j(t) \\
    S_2  (t) =& \frac16  \sum_{j =1}^{12} \sin ( j \pi/3) \delta \theta_j(t) 
\end{align}
The deformations $S_1$ and $S_2$ over time are plotted in Fig.~\ref{Fig2}d. 
Geometrically, $S_1$ and $S_2$ correspond to independent shearing deformations of the area enclosed by the ring. We then define $A(t) = \sqrt{ S_1^2(t) + S_2^2(t)}  $ and $\varphi(t) = \arctan(S_2/S_1)$ to be the polar coordinates (i.e. amplitude and angle) in Fig.~\ref{Fig2}d. 
The flux is defined to be
\begin{align}
    \Psi(t_c) = \dot \varphi e^{ \dot A t_c / A } \label{eq:flux} 
\end{align}
where $\dot \varphi $ and $\dot A/A $ are averaged over all data points between the release from the wall and the end of the recorded data (which corresponds to the projectile exiting the field of view, or deformations large enough that self intersection between the vertices occurs). Here $t_c=1$~s is a characteristic time scale for the experiment. Notice that the flux quantity captures the unidirectionality of the wave via the prefactor $\dot \varphi$ as well as the amplification or attenuation via the exponent $\dot A /A$. In Fig.~\ref{Fig2}e, data points denote weighted averages of all experiments and the error bars indicate standard deviation uncertainties arising from the particle tracking resolution and linear fits of the growth rate and phase velocity.

\subsubsection{Center of mass motion} 
To capture the center of mass motion, we model the projectile as a square of side length $R$ and we assume its only modes of deformation are the two shearing modes $S_1=u_{xx}-u_{yy}$ and $S_2=2u_{xy}$, where $u_{ij}$ is the strain tensor (assumed to be uniform throughout the square). The force on the center of mass of the block is proportional to the shear stress along the bottom surface. Hence we have:
\begin{align}
    \mqty( F_x \\ F_y  ) = \alpha_1 \mqty( \sigma_{xy} \\  \sigma_{yy} )  = \mqty(\sigma_2 \\ -\sigma_1 ) \label{eq:model1} 
\end{align}
where $\alpha_1$ is a proportionality constant and $\sigma_1=\sigma_{yy} -\sigma_{xx}$ and $\sigma_2 = 2\sigma_{xy}$ are the shear stresses corresponding to the stress tensor $\sigma_{ij} $. The coupling $\kappa^a$ induces an asymmetric shear coupling parameterized by an overall magnitude $\alpha_2$ and angle $\theta$
\begin{align}
    \mqty( \sigma_1 \\ \sigma_2 )  = \alpha_2 \mqty( \cos \theta & \sin \theta \\ -\sin \theta & \cos \theta  ) \mqty(S_1 \\ S_2 ) \label{eq:model2} 
\end{align}
If we denote the center of mass of the block by $(x,y)$, then we have $S_1 = - 2y/R$ and $S_2 = 2x/R$. Thus we obtain the equation of motion:
\begin{align}
     \mqty( \ddot x \\ \ddot y) = \frac{2\alpha_1 \alpha_2}{LM} \mqty( \cos \theta & - \sin \theta \\ \sin \theta & \cos \theta  ) \mqty( x \\ y ) \label{eq:modelmotion} 
\end{align}
where $M$ is the mass of the block.  

In Fig.~\ref{Fig2}f, we plot the ratios $p_{y, \text{out}}/p_{y, \text{in} }$ and $p_{x, \text{out}}/p_{y, \text{in} }$  for the simple box model. These quantities are defined as follows. We integrate Eq.~(\ref{eq:modelmotion}) using as initial conditions $(x,y) =0$ and $\dot y = p_{y,\text{in}}/(M \Sigma ) $. Here $\Sigma$ is a prefactor that sets the inelastic loss in the experiment that occurs upon impact.
We then terminate the integration when $y = 0$ (meaning that that block is departing the floor) and define the outgoing momentum $(p_{x,\text{out}}, p_{y,\text{out}})$. 
Since Eq.~(\ref{eq:modelmotion}) is linear, the quantities $(p_{x,\text{out}}/p_{y,\text{in}}, p_{y,\text{out}}/p_{y,\text{in}} )$ 
are independent of the numerical prefactors and only depend on the value of $\theta$ and $\Sigma$. 
The value of 
$(p_{x,\text{out}}/p_{y,\text{in}}, p_{y,\text{out}}/p_{y,\text{in}} )$
is easily determined numerically for all $\theta \in [-\pi/2,\pi/2]$. For $\theta \ll 1$, one can also write the leading order expression:
\begin{align}
    \frac{p_{y,\text{out}}}{p_{y,\text{in} }} =& \Sigma  +\order{\theta^2}\\
    \frac{p_{x,\text{out}} }{p_{y,\text{in}}} =& \Sigma  \theta  +\order{\theta^2} 
\end{align}
To compare to experiments, the value $\Sigma$ can be determined by setting the ratio of $p_{y,\text{out}}/p_{y,\text{in}}$ at $\kappa^a/\kappa =0$. Finally, one needs to determine the relationship between $\theta$ and $\kappa^a/\kappa$. Notice that the eigen-frequencies of Eq.~(\ref{eq:modelmotion}) are:
\begin{align}
    \omega^2 \propto 1+  i \tan \theta
\end{align}
Similarly, ignoring dissipation, the eigenvalues of the continuum model take the form:
\begin{align}
    \tilde \omega^2 \propto 1 + i \frac{8 \pi \kappa^a}{N \kappa }
\end{align}
Since one cycle in shear space should correspond to one cycle in the vibration of the chain, we  equate $\Re(\omega)/\Im(\omega) = \Re(\tilde \omega)/\Im(\tilde \omega)$ and obtain 
\begin{align}
    \theta = \arctan(\frac{8 \pi }{N} \frac{\kappa^a}{\kappa})
\end{align}
The box model only introduces  one additional parameter $\Sigma$, which sets the overall inelastic loss upon impact but before rebound. By examining the rebound at $\kappa^a/\kappa=0$, we find $\Sigma =0.42$. The theoretical predictions are compared against experimental measurements in Fig.~\ref{Fig2}k. Here, the markers indicate averages over all runs of the experiment and the error-bars indicate the full range.

\subsection{Impact of an odd wall}

\subsubsection{Odd elasticity and response to a point force}

Here we derive the point response in the continuum limit of the wall. 
In the continuum, the internal forces are captured by 
the stress tensor $\sigma_{ij}$ and the geometric deformation is captured by the (linearized) strain tensor $u_{ij} = \frac12( \partial_i u_j + \partial_j u_i )$, where $u_i$ is the displacement field. For small deformations, the stress-strain relationship can be approximated as linear $\sigma_{ij} = C_{ijmn} u_{mn}$. The form of the elastic modulus tensor $C_{ijmn}$ can be deduced on the basis of symmetries and conservation laws. Since angular momentum is conserved (i.e. there are no external sources of angular momentum in the bulk of the system), we have $C_{ijkl}=C_{jikl}$. Moreover, since the internal forces are not triggered by solid body rotation, we have $C_{ijkl}=C_{ijlk}$.  
Since the microscopic structure has $C_3$ symmetry, the macroscopic elastic modulus tensor must be isotropic. These three conditions imply that the elastic modulus tensor takes the following form~\cite{Scheibner_NatPhys2020}:
\begin{align}
C_{ijkl} =&B\delta_{ij} \delta_{kl} +  \mu (\delta_{il} \delta_{jk} + \delta_{ik} \delta_{jl} - \delta_{ij} \delta_{kl} ) 
+ K^o E_{ijkl}  
\label{eq:el} 
\end{align}
where $E_{ijkl} = \frac12[ \epsilon_{il} \delta_{jk} + \epsilon_{ik} \delta_{jl} + \epsilon_{jl} \delta_{ik} + \epsilon_{jk} \delta_{il}]$ and where $\delta_{ij}$ and $\epsilon_{ij}$ are the Kronecker and Levi-Civita symbols, respectively. Here $B$ and $\mu$ are the well known bulk and shear moduli, respectively, while the parameter $K^o$ is an example of an \emph{odd elastic} modulus. In a pictoral notation, we may write the stress and strain in the following basis:
\begin{align}
\pressure =& \sigma_{xx}+\sigma_{yy} & \dilation =& u_{xx} +u_{yy} \\
\strone =& \sigma_{xx} - \sigma_{yy} & \uone=& u_{xx} - u_{yy} \\
\strtwo =& \sigma_{xy} + \sigma_{yx} & \utwo=& u_{xy}+u_{yx}
\end{align}
Here, positive $\uone$ is equal to negative $S_2$ and $\utwo$ is equal to positive $S_1$ in the notation of the main text.
Using this notation, Eq.~(\ref{eq:el}) may be written as:
\begin{align} 
\mqty( \pressure \\ \strone \\ \strtwo  ) = 2
\mqty( B & 0 & 0 \\ 0 & \mu & K^o \\ 0 & -K^o & \mu  )
\mqty( \dilation \\ \uone \\ \utwo )
\end{align} 
It is useful to cast the equations into an overall stiffness $E$, known as the Young's modulus, as well as two dimensionless parameters: the Poisson's ratio $\nu$ and the so-called odd ratio $\nu^o$.  These parameters are given by:
\begin{align}
E =& \frac{  4 B [\mu^2 + (K^o)^2]   }{ \mu(\mu+   B)+ (K^o)^2 } \to  4 \mu \qty[ 1 + \qty(\frac{K^o}{\mu} )^2 ]   \label{eq:e}  \\
\nu =& \frac{\mu(B-\mu) - (K^o)^2 }{ \mu (B+\mu) + (K^o)^2} \to 1\label{eq:nu} \\
\nu^o =& \frac{BK^o}{ \mu (B+\mu) +(K^o)^2} \to \frac{K^o}{\mu} \label{eq:nuo}
\end{align}
The honeycomb lattice is approximately incompressible (to linear order). Hence in Eq.~(\ref{eq:e}-\ref{eq:nuo}), we show the limiting expressions as $B \to \infty$.

To compute the response to a point force, we must solve the bulk equation for static equilibrium  $\partial_i \sigma_{ij} =0$. To do so, it is useful to introduce the airy stress function $\chi$ defined via $\sigma_{ij} = \epsilon_{il} \epsilon_{jk} \partial_l \partial_k \chi$. Then $\partial_i \sigma_{ij} =0$ becomes $\Delta^2 \chi =0$. We consider a point force of magnitude $F$ on the boundary of a semi-infinite medium with stress free boundaries. 
Then the airy stress function is given by~\cite{Landau7}:
\begin{align}
\chi = - \frac{F}{\pi} r \phi \cos \phi  
\end{align} 
where $\phi$ is the angle with respect to the direction of the applied force. The corresponding stress is: $\sigma_{rr}= -\frac{2F}{\pi r} \cos \phi$ and $\sigma_{r\phi} = \sigma_{\phi \phi}=0$. The strain is obtained by inverting Eq.~(\ref{eq:el}):
\begin{align}
u_{ij} = \frac{1}{E} \bigg\{ & (1-\nu) \delta_{ij} \delta_{mn} + (1+\nu) ( \delta_{im} \delta_{jn} + \delta_{in} \delta_{jm} - \delta_{ij} \delta_{mn} ) -  2 \nu^o  E_{ijmn}  \bigg\} \sigma_{mn} \label{eq:inv}
\end{align}
Eq.~(\ref{eq:inv}) yields the differential equations:
\begin{align}
 u_{rr}=& \partial_r u_r = -\frac{4F}{\pi E r } \cos \phi  \\
 u_{\phi \phi} =&  \frac1r \partial_\phi u_\phi + \frac{u_r}{r} =   \nu \frac{4F}{\pi E r } \cos \phi \\
2 u_{r \phi} =& \partial_r u_\phi - \frac{u_\phi}r + \frac1r \partial_\phi u_r = -\nu^o \frac{8F}{\pi E r} \cos \phi
\end{align}
The solution to these equations is given by:
\begin{align}
u_r =& \frac{4F}{\pi E}\qty[- \log r \cos \phi + \frac{\nu-1}2 \sin \phi - \nu^o (\sin \phi + \phi \cos \phi ) ]\\
u_\phi =& \frac{4F}{\pi E} \qty[ \frac{1 + \nu + 2 \log r}{2} \sin \phi +\frac{\nu-1}2 \phi \cos \phi + \nu^o \phi \sin \phi ] 
\end{align}
Upon setting $\nu=1$, as it is appropriate for the incompressible limit, we obtain the expressions 
\begin{align}
u_r =& \frac{4F}{\pi E}\qty[- \log r \cos \phi  - \nu^o (\sin \phi + \phi \cos \phi ) ]\\
u_\phi =& \frac{4F}{\pi E} \qty[ (1  +\log r ) \sin \phi + \nu^o \phi \sin \phi ] 
\end{align}
The the $y$-component of displacement is given by $u_y =- u_r \cos \phi + u_\phi \sin \phi $, yielding Eq.~(\ref{eq:uy}) of the main text.

\subsubsection{Rayleigh wave dispersion} \label{sec:Rayleigh} 

Here we derive the Rayleigh wave dispersion shown in Fig.~\ref{Fig3}e. 
The bulk equations of motion reads 
\begin{align}
    \rho \partial_t^2 u_j = \partial_i \sigma_{ij} \label{eq:Rwave}
\end{align}
We assume that the medium occupies the $y<0$ half plane, which implies the boundary conditions
\begin{align}
\sigma_{yi}(y=0)=0 \quad \text{and} \quad  u_i (y \to -\infty) = 0\label{eq:bndry}
\end{align}
where $u_i$ is the displacement field. 
The stress tensor $\sigma_{ij}$ is given by $\sigma_{ij} = C_{ijmn} \partial_m u_n$. It is useful to parameterize the elastic modulus tensor in terms of the Young's modulus $E$, Poisson's ratio $\nu$, and odd ratio $\nu^o$. We can then non-dimensionalize the equations of motion by defining $r \equiv k_y/k_x$ and $ \tqx = k_x \sqrt{\frac{E}{\rho \omega^2}} $, where $\vb k$ is the wave-vector. 
In this notation, Eq.~(\ref{eq:Rwave}) takes the form:

\begin{align}
\mqty[u_x \\ u_y ] = \underbrace{\frac{\tqx^2}{ 2(1-\nu )  } 
\mqty[ \frac{(r^2+1)(1-\nu^2)}{ (\nu+1)^2 +4 (\nu^o)^2  } +1
 & \frac{2(r^2+1)(1-\nu) \nu^o}{ (\nu+1)^2 +4 (\nu^o)^2  } +r  \\
-\frac{2(r^2+1)(1-\nu) \nu^o}{ (\nu+1)^2 +4 (\nu^o)^2  } +r & \frac{(r^2+1)(1-\nu^2)}{ (\nu+1)^2 +4 (\nu^o)^2  } +r^2
] }_{M(r,\tqx) }
\mqty[u_x \\ u_y ]
\end{align}

Hence, we obtain a secular equation 
\begin{align}
    0=\det[ M (r, \tqx) -1]  \label{eq:poly1}
\end{align}
which is equivalent to 
\begin{align}
    0 = (r^2+1)^2 \tqx^4 -  [ 2(1- \nu^2 )+  (\nu+1)^2 +4 (\nu^o)^2   ] (r^2+1) \tqx^2 +2 (1-\nu) [(\nu+1)^2 +4 (\nu^o)^2 ]
\end{align}
The solutions to Eq.~(\ref{eq:poly1}) take the form:
\begin{align}
    r_1^\pm =& \pm \sqrt{ \frac{ 2(1-\nu^2) +  (\nu+1)^2 +4 (\nu^o)^2 - \sqrt{ (1+\nu)^4 -8( 1 -6 \nu + \nu^2 )(\nu^o)^2 +16 (\nu^o)^4 }   }{2 \tqx^2 }    -1} \\
    r_2^\pm =& \pm  \sqrt{ \frac{ 2(1-\nu^2) +  (\nu+1)^2 +4 (\nu^o)^2 + \sqrt{ (1+\nu)^4 -8( 1 -6 \nu + \nu^2 )(\nu^o)^2 +16 (\nu^o)^4 }   }{2 \tqx^2 }    -1}
\end{align}
Each solution comes with a normalized vector $n_i (r, \tqx) \in \ker [M (r, \tqx)-1 ] $. Each $n_i$ represents a candidate mode for a Rayleigh wave. We are looking for linear combinations of candidate modes that satisfy Eq.~(\ref{eq:bndry}). Moreover, we require $\Im(r \tqx) <0$ for all modes in the superposition in order to ensure that each mode decays into the bulk. The condition $\Im(r \tqx) <0$ implies that linear combinations are of the form:  
\begin{align}
n_i^{++}(\tqx )  \equiv&  a_1 n_i(r^+_1, \tqx) + a_2 n_i(r^+_2, \tqx)  \\
n_i^{+-}(\tqx )  \equiv&  a_1 n_i(r^+_1, \tqx) + a_2 n_i(r^-_2, \tqx)  \\
n_i^{-+}(\tqx )  \equiv&  a_1 n_i(r^-_1, \tqx) + a_2 n_i(r^+_2, \tqx)  \\
n_i^{--}(\tqx )  \equiv&  a_1 n_i(r^-_1, \tqx) + a_2 n_i(r^-_2, \tqx)  
\end{align}
We can express the stress at the boundary by:
\begin{align}
    \mqty[\sigma_{yx}  \\ \sigma_{yy} ] = \mqty[ S_{x1}^{\pm \pm } (\tqx) & S_{x2}^{\pm \pm } (\tqx)  \\ S_{y1}^{\pm \pm } (\tqx) & S_{y2}^{\pm \pm } (\tqx )   ]   \mqty[ a_1 \\ a_2 ]
\end{align}
where we have introduced the matrix $S^{\pm \pm} (\tqx)$ that is determined by computing $\sigma_{yj} = C_{ijmn} k_i k_m n_n^{\pm \pm }  $. Finally, the stress free boundary condition Eq.~(\ref{eq:bndry}) can be expressed algebraically as:
\begin{align}
   0= g^{\pm \pm }(\tqx) \equiv \det[S^{\pm \pm } (\tqx)]
\end{align}

For the special case that $\nu=1$, which corresponds to an incompressible solid, the algebraic expressions may be written explicitly. One finds:
\begin{align}
    r_1^\pm =& \pm i \\
    r_2^\pm =& \pm  \sqrt\frac{4 + 4 (\nu^o)^2 - \tqx^2 }{\tqx^2 }
\end{align}
Moreover, $g^{\pm \pm }$ up to an unimportant denominator reads
\begin{align}
    g^{++}=&0 \implies 0=4[1+(\nu^o)^2] +\tqx^4 - 2 ( i \nu^o +2 )\tqx^2 - (2 i \nu^o \tqx +\tqx^3) \sqrt{\tqx^2 - 4[1+(\nu^o)^2]}\\
    g^{+-}=&0  \implies 0= 4[1+(\nu^o)^2] +\tqx^4 - 2 ( i \nu^o +2 )\tqx^2 + (2 i \nu^o \tqx + \tqx^3) \sqrt{\tqx^2 - 4[1+(\nu^o)^2]} \\
    g^{--}=&0 \implies  0=4[1+(\nu^o)^2] +\tqx^4 + 2 ( i \nu^o -2 )\tqx^2 + (2 i \nu^o \tqx - \tqx^3) \sqrt{\tqx^2 - 4[1+(\nu^o)^2]} \\ 
    g^{-+}=&0 \implies  0=4[1+(\nu^o)^2] +\tqx^4 + 2 ( i \nu^o -2 )\tqx^2 - (2 i \nu^o \tqx - \tqx^3) \sqrt{\tqx^2 - 4[1+(\nu^o)^2]}
\end{align}
Notice that $g^{++}=0$ and $g^{+-}=0$ are both equivalent to the condition
\begin{align}
    \tqx^6 - (6-4 i \nu^o ) \tqx^4 +4 (1 - i \nu^o)( 2 - i\nu^o ) \tqx^2 - 4 (1 - i \nu^o)^2 =0 \label{eq:sec1}
\end{align}
while $g^{--}=0$ and $g^{-+}$ are both equivalent to the condition
\begin{align}
    \tqx^6 - (6+4 i \nu^o ) \tqx^4 +4 (1 + i \nu^o)( 2 +i\nu^o ) \tqx^2 - 4 (1 + i \nu^o)^2 =0 \label{eq:sec2}
\end{align}

Inspection of the roots of Eq.~(\ref{eq:sec1}) and Eq.~(\ref{eq:sec2}) along with the condition $\Im(r \tqx) < 0$ reveals that the roots of Eq.~(\ref{eq:sec1}) should be taken for right traveling waves ($\Re(\tqx) >0$) and the roots of Eq.~(\ref{eq:sec2}) should be taken for left traveling waves ($\Re(\tqy) < 0$). In the main text, we define $q_x=-\Im(k_x)$.

\subsubsection{Experimental strains and vibrations}
In Fig.~\ref{Fig3}, we measure the local displacement field $u_i$ of each node. The colormap overlaying the reconstruction in Panels a and c shows $u_y$ averaged over each hexagon. The space-time heat maps show the average value of $u_y$ along averaged along vertical slices. For Panel d, we compute the strain tensor $u_{ij} =\frac12 ( \partial_i u_j + \partial_j u_i)$ from the displacement field and average over all space. Shear 1 is defined as $S_1=(u_{xy}+u_{yx})/2$ and Shear 2 is defined as $S_2=(u_{yy}-u_{xx})/2$.

In Panel e, the center of vibration $\ell$ is defined as 
\begin{align} 
\ell = \frac{ \langle x \, \hat V_{t} [u_y(\vb x,t)] \rangle_{\vb x} }{ \langle \hat V_{t} [u_y (\vb x, t)] \rangle_{\vb x} }  \label{eq:ell}
\end{align} 
where the variance $\hat V$ is over time and the average is over space. In these experiments, the value of the passive bond stiffness is taken at the calibrated value $\kappa = 48$~mN~m/rad, see Section~S2.

\textbf{Acknowledgments.} \\
 We thank Daan Giesen, Tjeerd Weijers, Ronald Kortekaas and Kasper van Nieuwland for technical assistance and Michel Fruchart for valuable discussions. M.B. acknowledges funding from the Fonds de la Recherche Scientifique-FNRS. C.C. and J.V. acknowledge funding from the European Research Council under Grant Agreement No.~852587. C.S. and V.V.~acknowledge support from the Simons Foundation, the Complex Dynamics and Systems Program of the Army Research Office under grant W911NF-19-1-0268, and the University of Chicago Materials Research Science and Engineering Center, which is funded by the National Science Foundation under Award No.~DMR-2011854. C.S.~acknowledges funding from the National Science Foundation Graduate Research Fellowship under Grant No.~1746045. 

\textbf{Author contributions.} \\
 M.B., C.S., V.V. and C.C. designed the research. M.B., J.V. and C.C. designed the experiment and performed the measurements. M.B., C. S. and J.V. analyzed results and performed numerical simulations. C.S. and V.V. carried out the theoretical calculations. M.B., C.S., J.V., V.V. and C.C. wrote the paper.

\textbf{Code and data availability.} \\
The data and analysis codes are available at 
https://doi.org/10.5281/zenodo.5502579

\textbf{Supplementary Information.}\\
Supplementary Information is available for this paper.

\clearpage

\begin{center}
\Large
{\bf Supplementary Information for: \\
Limit cycles turn active matter into robots
}
\end{center}

\newcommand{\epsilonb}{\boldsymbol{\epsilon}}

\renewcommand{\theequation}{S\arabic{equation}}
\setcounter{equation}{0}
\setcounter{figure}{0}
\setcounter{section}{0}
\renewcommand{\thetable}{S\arabic{table}}  
\renewcommand{\thefigure}{S\arabic{figure}}
\renewcommand\figurename{Fig.}
\renewcommand{\thesection}{S\arabic{section}}

\section{Mathematical derivations}

\subsection{Wave equation for 1D Chain} \label{sec:1DChain}

In this section, we derive the equations of motion for a 1D chain of motorized vertices. We consider a 1D chain of point particles of mass $m=1$ located at positions $\vb x_\alpha$, $\alpha =0 ,1 , \dots , N-1$.  We assume each bond has a rest length $a$, and we denote the angular tension in each vertex $\alpha$ by $\tau_\alpha$. Additionally, we will at first assume that the bonds are extensible and have a Hookean spring constant $k$. (We will subsequently take $k \to \infty$ to obtain inextensibility constraints.)
The equations of motion are constructed via Newton's laws:
\begin{align}
\ddot {\vb x}_\alpha =& \vb F (\underline {\vb x}) \\
=& \epsilonb \cdot \qty[ \frac{    \vb x_{\alpha +1} - \vb x_{\alpha } }{ \abs{\vb x_{\alpha +1} - \vb x_{\alpha }}^2 } ( \tau_\alpha (\underline{\vb x}) - \tau_{\alpha +1} (\underline{\vb x})  ) +  \frac{  \vb x_{\alpha -1} - \vb x_{\alpha } }{ \abs{\vb x_{\alpha -1} - \vb x_{\alpha }}^2 } ( \tau_\alpha (\underline{\vb x}) - \tau_{\alpha -1} (\underline{\vb x})  ) ]  \nonumber \\
&+ k \qty[  \frac{ \vb x_{\alpha+1} - \vb x_{\alpha }  }{\abs{\vb x_{\alpha +1} - \vb x_\alpha } } ( \abs{\vb x_{\alpha +1} - \vb x_\alpha } -a  )+ \frac{ \vb x_{\alpha-1} - \vb x_{\alpha }  }{\abs{\vb x_{\alpha -1} - \vb x_\alpha } } ( \abs{\vb x_{\alpha -1} - \vb x_\alpha } -a ) ] \label{eq:EOM}
\end{align}
where $\epsilonb$ is the Levi-Civita symbol. The first term in Eq.~(\ref{eq:EOM}) are the forces due to the angular tensions, while the second term is the Hookean resistance to stretching. 
The functions $\tau_\alpha(\underline{\vb x})$ determine the vertex bending properties. For the experiments, we model the vertices via
\begin{align}
    \tau_\alpha (\vb x) = \kappa ( \theta_\alpha - \theta^0) + \kappa^a ( \theta_{\alpha+1} - \theta_{\alpha-1}  ) \label{eq:tau}
\end{align}
Here,  $\theta^0$ is the rest angle (assumed to be identical for all vertices) and $\theta_\alpha$ is the angle of the vertex at point $\alpha$:
\begin{align}
    \theta_\alpha = \arccos( \frac{ (\vb x_{\alpha+1} - \vb x_\alpha )\cdot ( \vb x_{\alpha-1} - \vb x_{\alpha}  )  }{\abs{\vb x_{\alpha+1} - \vb x_\alpha } \abs{\vb x_{\alpha-1} - \vb x_{\alpha} } }  ) \label{eq:theta} 
\end{align}
Equations~(\ref{eq:EOM}, \ref{eq:tau}, \ref{eq:theta}) constitute a closed system of equations.

Next, we take the continuum limit of Eq.~(\ref{eq:EOM}). We let $\vb x$ become a function of a continuous parameter $\vb x(\lambda)$ where $\lambda \in [0, a N]$ and approximate the differences by a Taylor expansion:
\begin{align}
\vb x_{\alpha +1} - \vb x_{\alpha} = a \vb x'(\alpha) + \frac{a^2} 2  \vb x''(\alpha) + \dots 
\end{align}
Expanding the discrete equations of motion to order $a^3$ yields
\begin{align}
    \ddot {\vb x} =&  -\frac{  \epsilonb \cdot \vb x '}{ \abs{\vb x'}^2}  \bigg\{ a^2 \kappa  \qty[ \frac{\vb x' \times \vb x^{(4)} }{\abs{\vb x'}^2} + 3 \frac{ \vb x '' \times \vb x ''' }{\abs{\vb x '}^2} +  2\frac{  ( \vb x'' \times \vb x')^3}{ \abs{\vb x'}^6}   + 6\frac{ (\vb x ' \cdot \vb x '')   [  (\vb x' \cdot \vb x'') (\vb x ' \times \vb x'') - \vb x ' \times \vb x''' ] }{\abs{\vb x'}^8} ]   \nonumber \\
    &  +2 a^3 \kappa^a \bigg[ \frac{ \vb x ' \times \vb x^{(5)} }{ \abs{ \vb x '}^2 }  
    -4 \frac{ ( \vb x' \cdot \vb x '')( \vb x' \times \vb x^{(4)})+( \vb x' \times \vb x'' )( \vb x' \cdot \vb x^{(4) }) }{\abs{ \vb x' }^4 } - 6 \frac{ (\vb x' \times \vb x ''') (\vb x ' \cdot \vb x ''')  }{ \abs{\vb x'}^4 } \nonumber \\
    &+12 \frac{ [(\vb x ' \cdot \vb x '')^2 - (\vb x' \times \vb x'')^2](\vb x' \times \vb x ''') +2 (\vb x' \cdot \vb x''')(\vb x' \cdot \vb x'') (\vb x' \times \vb x'')  }{ \abs{\vb x'}^6 }  \nonumber \\
    &-24 \frac{ (\vb x' \times \vb x'')(\vb x' \cdot \vb x '')[ (\vb x ' \cdot \vb x'')^2-(\vb x' \times \vb x'')^2 ] }{\abs{\vb x'}^8}  \bigg]   \bigg \} \nonumber \\  
    &-\frac{ \epsilonb \cdot \vb x '' \abs{\vb x'}^2 - 2 \epsilonb \cdot \vb x' (\vb x' \cdot \vb x'') }{ \abs{\vb x'}^4 } \bigg\{ a^2 \kappa \qty[ \frac{ \vb x' \times \vb x '''}{ \abs{\vb x '}^2} - 2 \frac{ (\vb x' \times \vb x'')( \vb x' \cdot \vb x'') }{\abs{\vb x'}^4} ]  \nonumber \\
    & + 2 a^3 \kappa^a \qty[  \frac{\vb x' \times \vb x^{(4)} }{\abs{\vb x'}^2} + 3 \frac{ \vb x '' \times \vb x ''' }{\abs{\vb x '}^2} +  2\frac{  ( \vb x'' \times \vb x')^3}{ \abs{\vb x'}^6}   + 6\frac{ (\vb x ' \cdot \vb x '')   [  (\vb x' \cdot \vb x'') (\vb x ' \times \vb x'') - \vb x ' \times \vb x''' ] }{\abs{\vb x'}^8} ]    \bigg\} \nonumber \\
    &+k a^2 \qty[  \frac{\vb x '  (\vb x' \cdot \vb x'')}{ \abs{\vb x'}^2 }- \frac{  \vb x''  }{ \abs{\vb x'}^3  } (1- \abs{\vb x '}) ]   + \order{a^4} \label{eq:long}
\end{align}

To simplify the equations of motion, we take $k \to \infty$, which imposes the constrain equation $\abs{\vb x '} =1$. In this case, we obtain:
\begin{align}
    \ddot {\vb x} =&  -  \epsilonb \cdot \vb x '  \bigg\{  a^2 \kappa  \qty[ \vb x' \times \vb x^{(4)}  + 3 \vb x '' \times \vb x '''  +  2  ( \vb x'' \times \vb x')^3  ]  + 2 a^3 \kappa^a \bigg[  \vb x ' \times \vb x^{(5)} -4  ( \vb x' \times \vb x'' )( \vb x' \cdot \vb x^{(4) }) \nonumber \\
    &-12 (\vb x' \times \vb x'')^2(\vb x' \times \vb x ''')    \bigg]   \bigg \} \nonumber \\
    &  - \epsilonb \cdot \vb x '' \bigg\{ a^2 \kappa (\vb x' \times \vb x''') + 2 a^3 \kappa^a \qty[\vb x' \times \vb x^{(4)}  + 3 \vb x '' \times \vb x '''  +  2  ( \vb x'' \times \vb x')^3  ]    \bigg\}   + \dv{\lambda}  (\vb x' T ) \label{eq:simp}
\end{align}
where $T (\lambda)$ is a tension field that enforces the constraint $\abs{\vb x'} =1$. (The field $T$ is analogous to the pressure field $p$ in an incompressible fluid). 
In principle, one must solve for $T(\lambda)$ to produce a nonlocal equation entirely in terms of ${\vb x}$ and its derivatives. For example, let us denote the coefficient of $\epsilonb \cdot \vb x'$ in Eq.~(\ref{eq:simp}) as $g(\lambda)$ and likewise the coefficient of $\vb x''$ as $f(\lambda)$. Then the condition $\abs{\vb x'}^2=1$ implies 
\begin{align}
0=&\abs{\dot {\vb x}' }^2 + \ddot {\vb x}' \cdot \vb x'  \label{eq:const1} 
\end{align}
When Eq.~(\ref{eq:const1}) is combined with Eq.~(\ref{eq:simp}), we obtain a differential equation for  $T$:
\begin{align}
  \abs{\dot {\vb x}'}^2 - ({\vb x }' \times \vb x '') (g+f') - (\vb x' \times \vb x''')f + (\vb x ' \times \vb x'')^2 T+ T '' =0 \label{eq:diffy}
\end{align}
In principle, we can represent the solutions of Eq.~(\ref{eq:diffy}) as a non-local kernel $T(\dot {\vb x}' , \vb x ', \vb x'', \dots)$ and place the nonlocal kernel into Eq.~(\ref{eq:simp}) to obtain a closed equation of motion. 
We note that in deriving Eq.~(\ref{eq:long}) we used 
\begin{align}
\theta_i = \sqrt 2 \sqrt{1- \frac{ (\vb x_{\alpha+1} - \vb x_\alpha )\cdot ( \vb x_{\alpha-1} - \vb x_{\alpha}  )  }{\abs{\vb x_{\alpha+1} - \vb x_\alpha } \abs{\vb x_{\alpha-1} - \vb x_{\alpha} } }  }
\end{align}
instead Eq.~(\ref{eq:theta}) in order to simplify the expansion. Moreover, for simplicity, we considered the dissipation free case in the derivation above, but the role of hinge dissipation $\Gamma$ may be added in a similar fashion.

\subsubsection{Geometry 1: straight line}
Since we have the full nonlinear equations, we may linearize them around various background geometries. We will do this in two cases: an infinite straight line and a circle. 
For the straight line, we write:
\begin{align}
\vb x(\lambda) = \mqty( \lambda \\ h(\lambda) )
\end{align}
We note that $\abs{\vb x ' } = 1 + \order{h^2}$, so the ansatz is consistent with the inextensibility condition. The linearization of Eq.~(\ref{eq:simp}) yields:
\begin{align}
    \ddot h  = a^2 \kappa h^{(4)} +2 a^3\kappa^a h^{(5)} 
\end{align}
as written in Eq.~(\ref{eq:wave}). The spectrum is 
\begin{align}
    \omega^2 =  a^2 \kappa q^4 + 2 i a^3 \kappa^a q^5
\end{align}
which has positive and negative imaginary branches. We note that Eq.~(\ref{eq:diffy}) is satisfied by $T=0$, so the tension does not enter the straight-line problem to leading order $h$.

\begin{figure}
    \centering
    \includegraphics{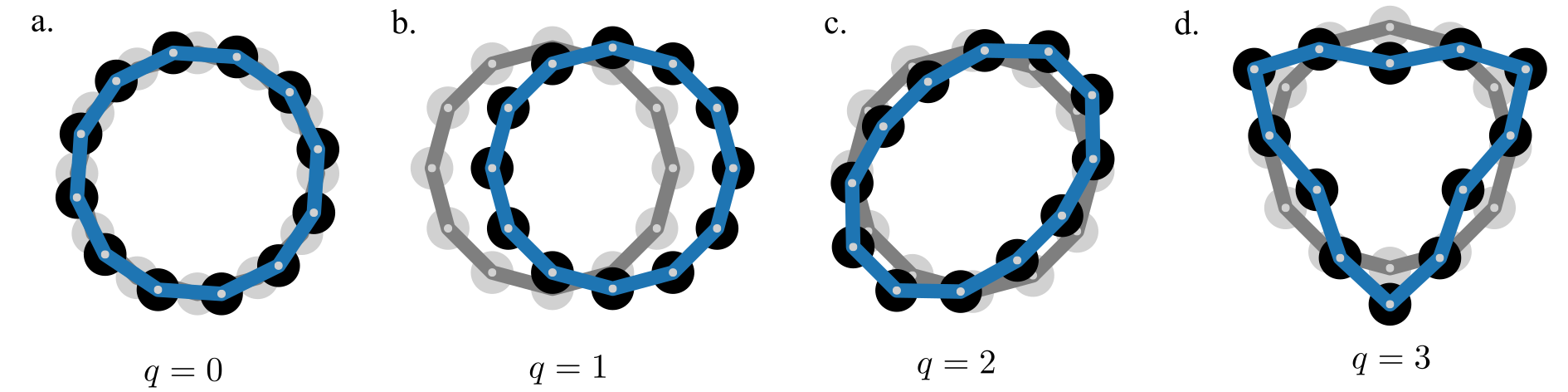}
    \caption{An illustration of the $q=0,1,2,3$ modes on a circular chain. Notice that $q=0$ is a solid body rotation and $q=1$ is a solid body translation, and hence are exact zero modes of the system. The $q=2$ mode is the lowest order deformation.}
    \label{fig:mode_def}
\end{figure}

\subsubsection{Geometry 2: circle} 
Next, we perform a linearization about a circular geometry. For simplicity of notation, we will measure all distances in terms of the radius of the circle $R$, e.g. $\vb x \to \vb x /R$ and $\lambda \to \lambda /R$. Without loss of generality, we may write $\vb x$ as:
\begin{align}
    \vb x = \mqty( \cos \lambda  \\ \sin \lambda   ) [1 + h(\lambda) ] +\mqty( -\sin  \lambda \\ \cos \lambda  ) g(\lambda)  
\end{align}
To leading order in $g$ and $h$, requiring $\abs{\vb x'} = 1 $  yields the requirement $h = - g'$. Thus we write 
\begin{align}
\vb x = \vb r(\lambda)  [1 -g'(\lambda) ] + \boldsymbol{\phi}(\lambda)  g(\lambda)
\end{align}
where $\vb r(\lambda) = (\cos  \lambda , \sin  \lambda  ) $ and $\boldsymbol{\phi} = \vb r' $. The equations of motion become:
\begin{align}
    \ddot g =& \boldsymbol{\phi} \cdot \ddot{\vb x} = A_\phi + T' \label{eq:g} \\
    -\ddot g' =& \vb r \cdot \ddot{ \vb x} = A_r - T  \label{eq:gd}
\end{align}
where
\begin{align}
    A_r =& 12 \kappa^a g'' - \kappa g ''' + 10 \kappa^a g^{(4)} - \kappa g^{(5)} -2 \kappa^a g^{(6)} \\
    A_\phi =& - \kappa g'' -2 \kappa^a g''' - \kappa g^{(4)} - 2 \kappa^a g^{(5)}
\end{align}
We can use  Eqs.~(\ref{eq:g}-\ref{eq:gd}) to eliminate the tension, and we obtain:
\begin{align}
    (1- \partial_\lambda^2) \ddot g = A_r' + A_\phi  = - a^2 \kappa g'' + 10 a^3 \kappa^a g''' - 2 a^2 \kappa g^{(4)} + 8 a^3 \kappa^a g^{(5)} - a^2 \kappa g^{(6)} - 2 a^3 \kappa^a g^{(7)}
\end{align}
Hence, the spectrum is given by:
\begin{align}
    \omega^2 =  a^2 \frac{ \kappa (q^2-2q^4+q^6) - a i \kappa^a ( 10  q^3 - 8  q^5 - 2  q^7)  }{1+q^2}
\end{align}
where $q$ takes integer values since $g(\lambda)$ must be periodic under $\lambda \to \lambda + 2\pi $. As shown in Fig.~\ref{fig:mode_def}, $q=0$ corresponds to a solid body rotation and $q=1$ corresponds to a solid body translation. Hence we obtain $\omega^2=0$ for $q=0$ and $q=1$.  The $q=2$ is the lowest geometric deformation of the object, and we obtain a finite-frequency response. 
Notice that the expression for $q=2$ matches exactly that of the straight chain.
Finally, for $q \gg 1$, we recover the result for the 1D chain $\omega^2 = a^2 \kappa q^4 + 2 i a^3 \kappa^a q^5$, as expected.

\subsection{Non-conservative forces and Hodge decomposition} 
Here we review how the notions of a nonconservative work cycle generalize to more arbitrary degrees of freedom. 
First we consider a system with $N$ discrete degrees of freedom, which we can represent as a manifold $\MM$ of dimension $N$.
Recall that the tangent space of $T_p \MM$ at a point $p \in M$ is a vector space of all derivatives evaluated at point $p$. The cotangent space $T_p^*\MM$ is a vector space consisting of all linear maps from $T_p \MM$ to the real line. Given a local set of coordinates $(\theta^1, \dots, \theta^N)$ on $M$ in a neighborhood containing $p$, we may use the coordinate derivatives $\{\pdv{\theta^1}, \dots,  \pdv{\theta^N} \}$ as a basis for $T_p\MM$ and define a basis $\{ \dd \theta^1, \dots, \dd \theta^N \}$  for $T^*_p\MM$ via the relationship $\dd \theta^\alpha [ \pdv{\theta^\beta}] = \tensor{\delta}{^\alpha_\beta}$. In this basis, a given covector $\lambda \in T^*_p \MM$ may be identified through its coordinates $\lambda_\alpha$ is the basis expansion $\lambda = \lambda_\alpha \dd \theta^\alpha$. Likewise for vectors $\lambda \in T_p \MM$  may be written as $\lambda=\lambda^\alpha \pdv{\theta^\alpha}$. For mechanical systems, we are particularly interested in a distinguished vector field $\tau=\tau_\alpha \dd \theta^\alpha$ that represents the configuration-dependent forces. This force field is defined by the property that for any trajectory $\gamma : [t_i, t_f] \to \MM$, the quantity $P=\tau[\dot \gamma]$ represents the physical power exerted by the given forces. (Recall that $\eval{\dot \gamma}_t = \eval{\dv{\theta^\alpha}{t}}_t \pdv{\theta^\alpha} \in T_{\gamma(t) } \MM$.) The total work done over the trajectory $\gamma$ is given by:
\begin{align}
    W[\gamma] = \int_{t_i}^{t_f} P(t) \dd t = \int_\gamma \tau  
\end{align}
Vector fields such as $\tau$ can be classified via their behavior under integration. This classification scheme, known as Hodge decomposition, relies on two distinct notions of a derivative. 
Recall that an arbitrary differential $k$-form $\omega$ may be written as
\begin{align}
    \omega = \sum_{ \alpha_1 < \dots < \alpha_k } \omega_{\alpha_1 \dots \alpha_k } \dd \theta^{\alpha_1} \wedge \cdots \wedge \dd \theta^{\alpha_k}. 
\end{align}
where $\wedge$ is the wedge product. 
The coefficients $\omega_{\alpha_1 \dots \alpha_k}$ are function of the point $p \in \MM$. The exterior derivative $d$ is a map from $k$ forms to $k+1$ forms given by
\begin{align}
    \dd \omega = \sum_{ \alpha_1 < \dots < \alpha_k } \sum_\gamma \pdv{\omega_{\alpha_1 \dots \alpha_k }}{\theta^\gamma } \dd \theta^\gamma \wedge \dd \theta^{\alpha_1} \wedge \cdots \wedge \dd \theta^{\alpha_k}.
\end{align}
Next, the Hodge dual, denoted by $\star$, is a vector space isomorphism between differential $k$ forms and $N-k$ forms.  
Its definition is most succinctly stated on a single basis element:
\begin{align}
    &\star \dd \theta^{\alpha_1} \wedge \dots \wedge \dd \theta^{\alpha_k} = 
    \frac{\sqrt{g}}{(N-k)!} \tensor{\epsilon}{^{\alpha_1 \cdots \alpha_k}_{\beta_1 \dots \beta_{N-k}} } \dd \theta^{\beta_1} \wedge \cdots \wedge \dd \theta^{\beta_{N-k}}  \label{eq:hodges}
\end{align}
In Eq.~(\ref{eq:hodges}), the quantity $g$ is the determinant of the metric $g_{\alpha \beta}$ on $\MM$ and $\epsilon_{\gamma_1 \cdots \gamma_N }$ is the totally antisymmetric Levi-civita symbol (whose indices may be raised using the metric tensor). The metric $g_{\alpha \beta}$ does not have an inherent physical meaning in the current context\textemdash nonetheless, the decomposition theorem stated below applied unambiguously regardless of the (arbitrary) choice of metric. 
Notice that $\star$ is a duality in the sense that 
\begin{align}
\star \star = 
\begin{cases} 
1 & N \text{ odd} \\
(-1)^k & N \text{ even} 
\end{cases} 
\end{align}
We may use $\star$ to induce an inner product on $k$ forms on a closed manifold. Given $\omega, \lambda \in \Lambda^k$ and taking $\MM$ to be closed, we define:
\begin{align}
 \langle \omega, \lambda \rangle = \int_\MM \omega \wedge \star \lambda 
\end{align}
Finally, we use the Hodge start to introduce a second type of derivative, $\delta$, known as the \emph{codifferential}. Unlike $d$, which is a linear map from $k \to {k+1}$ forms, the codifferential is a map from ${k} \to {k-1}$ forms. The codifferential is explicitly given by $\delta = \star^{-1} \dd \star $, and it readily verified that $\langle \omega , \dd \lambda \rangle = \langle \delta \omega, \lambda \rangle$ for arbitrary $k$ form $\omega$ and $k-1$ form $\lambda$. Hence $\delta$ is the adjoint of $d$ with respect to the inner product $\langle \cdot , \cdot \rangle$. We note that $\dd$ and $\delta$ both have the property $\dd^2=\delta^2=0$. 

Finally, we are prepared to state the Hodge decomposition theorem. The simplest formulation of the Hodge decomposition theorem states that an arbitrary $k$-form $\tau$ on a closed manifold $\MM$ may be uniquely decomposed into three pieces:
\begin{align}
    \tau = \dd V + \delta A +h
\end{align}
Here, $V$ is a $k-1$ form, $A$ is a $k+1$ form, and $h$ is a $k$-form that satisfies the differential equation $\Delta h=0$, where $\Delta=\dd\delta + \delta\dd $ is the Laplace-deRham operator. Such $h$ are known as harmonic forms. One can show that $h$ obeys $\dd h= \delta h =0$. 
In the main text, we use a special case of the Hodge decomposition theorem in which $\tau$ is a one-form representing generalized forces. As alluded to before, the Hodge decomposition theorem allows us to classify vector fields $\tau$ based on their integral properties. Suppose that the trajectory $\gamma(t)$ forms a closed loop that is also contractible to a point (i.e. it is a topologically trivial curve in parameter space). In this case, we may write $\gamma = \partial \VV$. Then we have 
\begin{align}
    W = \int_\gamma \tau = \int_\VV \dd \tau = \int_\VV \dd \delta A  
\end{align}
where we used the fact that $\dd \dd V = \dd h =0$. However, if $\gamma$ enclosed a non-contractible loop (i.e. it encloses a whole in parameter space), then $h$ can contribute to the work done. In this sense $h$ can be interpreted as the force due to a multi-valued potential. In certain mechanical contexts, the degrees of freedom do not naturally form a closed manifold $\MM$, but instead form an open manifold. Certain generalization of the Hodge decomposition are possible under assumption about the force fall-off in more general contexts. For a continuous material, one can consider the internal degrees of freedom at each point in space (such as the strain field). The Hodge decomposition theorem can be applied to conjugate forces (e.g. the stresses) at each point in space. 

\clearpage 

\section{Calibrations} \label{sec:calibration}

\subsection{Calibration of robotic vertices}\label{sec:calel}
\begin{figure}
    \centering
    \includegraphics[width=1\columnwidth]{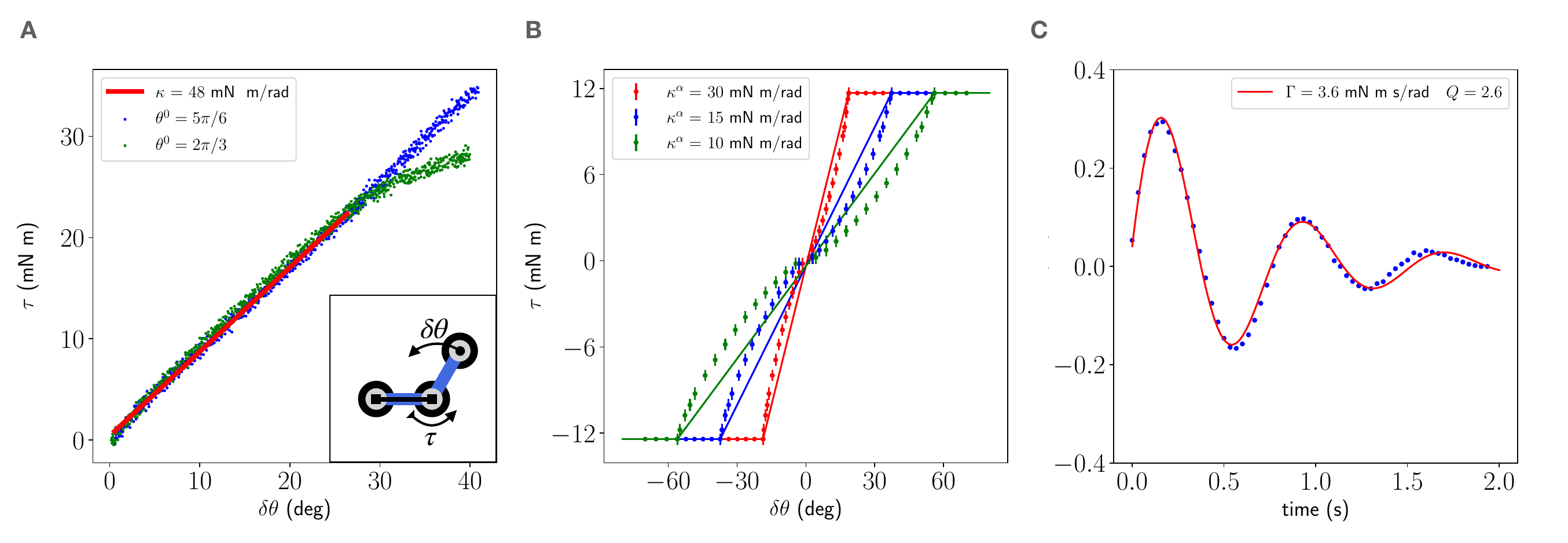}
    \caption{(\textbf{A})~The restoring torque $\tau$ measured over a range of twist angles $\delta \theta$. The data shown is for rest angle $\theta^0 = 5\pi/6$ (blue points) and $\theta^0 = 2 \pi/3$ (green points). The slope of the linear regression  yields a value $\kappa = 48$ mN m/rad. (\textbf{B})~For different values of the active feedback, the motorized torque is approximately linear up to saturation of the DC motor at $\tau_{max} \approx 12$ mN m. The parameter $\kappa^a$ is taken to be the slope of the line connecting the saturation points. (\textbf{C})~The decaying oscillations of the building block were fitted to the solution of a damped harmonic oscillator indicating a quality factor of $Q=2.6$.  }
    \label{fig:calib}
\end{figure}

The motorized building blocks described in Section~\ref{sec:construction} are modeled via an angular stiffness $\kappa$, active feedback $\kappa^{a}$, and angular dissipation $\Gamma$. As shown in the inset of Fig.~\ref{fig:calib}A, we determine $\kappa^a$ and $\kappa$ by placing a single robotic vertex in a rotational testing device (Instron E3000 linear-torsional) equipped with a torsional load cell (angular resolution: $2.5$ arcsecs; torque resolution: $1$ mN m). To determine the passive stiffness $\kappa$ provided by the rubber band, the torque was measured as a function of angular deviation. The value of $\kappa$ is taken to be the slope of the linear fit. We find $\kappa= 48$ mN m/rad for both the hexagonal ring with rest angle $\theta^0=2\pi/3$ and the twelve-sided projectile with rest angle $\theta^0=5\pi/6$.

To calibrate the torque provided by the motors, we remove the rubber band and measure the torque as a function of applied voltage, see Fig.~\ref{fig:calib}B. When implemented in the robotic devices, the applied voltage is proportional to the angular deviation $\delta \theta$ of the neighboring vertices. The programmed proportionality constant gives rise to different torque-angle curves. The value of $\kappa^a$ is taken to be the slope of the straight line connecting the saturation points of the motors. 
To determine $\Gamma$, we place the rubber band on the linkage system, and we apply an angular perturbation to the free edge. We measure the angular displacement and extract $\Gamma$ from the decay rate, see Fig.~\ref{fig:calib}C.  

\subsection{Calibration of odd ratio} \label{sec:calnu}
\begin{figure}
    \centering
    \includegraphics[width=1\columnwidth]{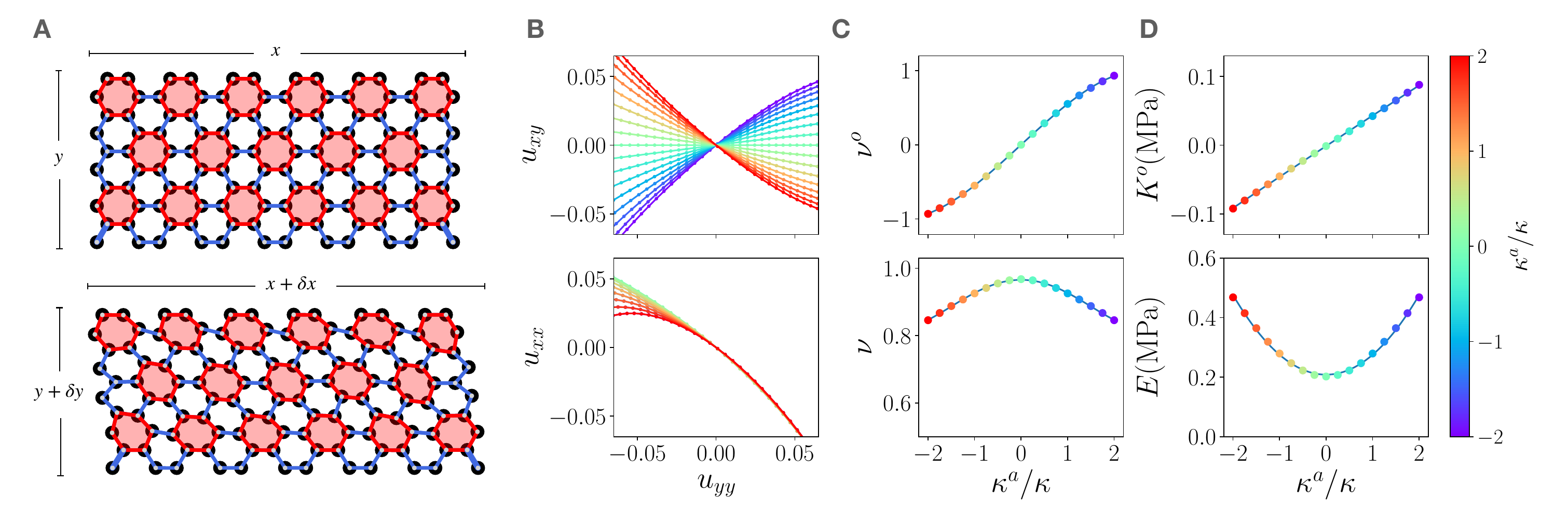}
    \caption{(\textbf{A})~Snapshots of a simulation of the active wall in its undeformed state (top) and after a quasistatic compression (bottom). The horizontal strains $u_{yy}=-\frac{\delta y}{y}$, the vertical strains $u_{xx}=-\frac{\delta x}{x}$ and the simple shear $u_{xy} =\frac{\delta x}{2y}$ are calculated by averaging the strains over all hexagons. The vertices within each red hexagon are coupled via active feedback $\kappa^a$.  (\textbf{B})~Shear strain $u_{xy}$ (top) and normal strain $u_{xx}$ (bottom) as a function of the normal strain $u_{yy}$ for a range of values of the ratio between the active feedback and the hinge stiffness $\kappa^a/\kappa$. These curves were fitted polynomially and the slope at $u_{yy}=0$ defines the odd ratio (Poisson's ratio). (\textbf{C})~The odd ratio (top) and Poisson's ratio (bottom) as a function of $\kappa^a/\kappa$. A cubic fit $\nu^0 = 0.58 \frac{\kappa^a}{\kappa}-0.03 \Big(\frac{\kappa^a}{\kappa}\Big)^3$ is used to calibrate the value of $\nu^o$ in experiments. (\textbf{D})~The odd modulus $K^o$ (top) and Young's modulus $E$ (bottom) as a function of $\kappa^a/\kappa$. }
    \label{fig:oddratio}
\end{figure}

To connect the ratio $\kappa^a/\kappa$ to the macroscopic odd ratio $\nu^o$ and odd modulus $K^o$ in Fig.~\ref{Fig3}, we perform numerical simulations of the 2D robotic medium. As depicted in Fig.~\ref{fig:oddratio}A, a simulated wall is subjected to a uniaxial compression. To reach a force balanced configuration, the vertices evolve according to a first order equation of motion. To apply the uniaxial compression, the $y$ (vertical) coordinate of the top and bottom row of particles is controlled, while the $x$ coordinate is allowed to evolve according to the internal forces, thereby creating sliding boundary conditions.  Upon compression, the strain data was sampled over all hexagons. When the tilt $u_{xy}$ is plotted as a function of $u_{yy}$, the odd ratio is given by $\nu^o = - \dv{u_{xy}}{u_{yy}}$ evaluated at $u_{yy}=0$ (Fig.~\ref{fig:oddratio}B). In Fig.~\ref{fig:oddratio}C, the odd and Poisson ratio are plotted as a function of $\kappa^a/\kappa$. To find the odd modulus $K^o = \frac{d \sigma_{xy}}{d u_{yy}}$, the transverse stress $\sigma_{xy}$ was calculated from force data of a compression simulation with top and bottom row particles constrained in both directions, and its slope evaluated at $u_{yy}=0$ (Fig.-\ref{fig:oddratio}D). The code for the simulations is uploaded on the public repository (see Main Text for link).

\begin{figure}
    \centering
    \includegraphics[width=0.5\textwidth]{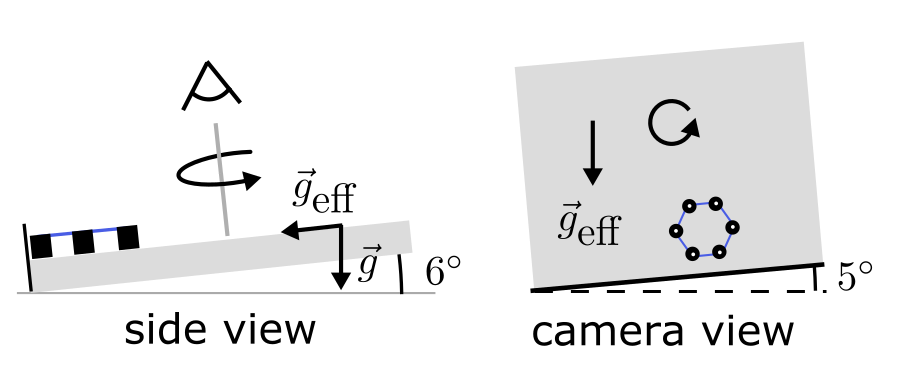}
    \caption{A schematic of the experimental setup for the air-table geometry for Fig.~\ref{Fig:wheel}g in the main text.}
    \label{fig:airtable}
\end{figure}

\clearpage 
\section{Description of Supplemental Movies}

\textbf{Summary Video.} 
Non-conservative forces and nonlinear work cycles endow an odd material with robotic functionalities. For three rigid linkages, odd interactions lead to asymmetric responses. When six odd vertices are connected in a hexagon, after an initial deformation, the internal non-conservative forces balance with dissipation and sustain a nonlinear work cycle. The nonlinear work cycle allows the hexagon to interact with its environment and power locomotion. Beyond cyclic deformation, the non-conservative forces control the impact of an odd projectile against a passive wall and, likewise, a passive projectile against an odd wall.

\medskip

\textbf{Supplemental Video 1.} 
Three rigid linkages are connected by motorized vertices. When a hand pushes in on the left, the right vertex contracts. On the contrary, when a hand pushes in on the right, the left vertex expands. See Fig. 1 in the main text for additional information.

\medskip

\textbf{Supplemental Video 2.}
Six odd vertices are connected in a hexagon, whose shape is summarized by three deformation modes: two shears $S_1$ and $S_2$, and a breathing mode $B$. When initially perturbed, the system evolves towards a limit cycle in the space of $S_1$ and $S_2$. Color indicates the phase angle $\varphi$. The nonlinear work cycle allows the odd wheel to interact with its environment and power locomotion. See Fig. 2 in the main text for additional information.

\medskip

\textbf{Supplemental Video 3.}~The collision of an odd projectile against a passive wall is controlled by shear coupling emerging from local non-conservative forces. See Fig. 3 in the main text for additional information. 

\medskip 

\textbf{Supplemental Video 4.}~The deformation of an odd wall under impact displays shear cycles emerging from the non-conservative feedback. See Fig. 4 in the main text for additional information.


\end{document}